\documentclass{pj}
\usepackage{graphicx}
\usepackage{psfrag} 
\usepackage{amssymb} 
\usepackage{amsmath}

\newcommand{\mb}[1]{\mathbf{#1}}
\newcommand{\mt}[1]{\mathcal{#1}}
\renewcommand\Re{\operatorname{Re}}

\begin{document}
\setcounter{page}{1}
\pjheader{}

\title[]{High-order functional derivatives of the diffracted field according to the permittivity-contrast function}
\footnote{\hskip-0.12in*\, Corresponding author:~Slimane~Arhab (slimane.arhab@univ-avignon.fr).}
\footnote{\hskip-0.12in\textsuperscript{1} Universit\'e d'Avignon et des Pays de Vaucluse, UMR 1114 EMMAH, 84018 Avignon Cedex, France.}

\author{Slimane~Arhab\textsuperscript{*, 1}, Dimitrios~Anagnostou\textsuperscript{1} and Maminirina~Joelson\textsuperscript{1}}

\runningauthor{}


\begin{abstract}
In this work, we propose to extend an approach to calculate at any order $(n)$, the functional derivative of the diffracted field with respect to the permittivity-contrast function. These derivatives obtained for different orders are used to perform an expansion of the data according to the studied model parameter. Its convergence is discussed throughout some numerical results, obtained in the case where the forward model used to simulate the diffracted field is built in the framework of the volume integral formulation. In particular, we show that taking into account higher order derivatives improve drastically the data-fitting. The numerical application considered consists of a cylindrical object illuminated by an incident field under a ${\rm TE}$ polarisation (electric component parallel to the invariance axis). 
\end{abstract}

\setlength {\abovedisplayskip} {6pt plus 3.0pt minus 4.0pt}
\setlength {\belowdisplayskip} {6pt plus 3.0pt minus 4.0pt}

\

\section{Introduction}
\label{introduction}
Reconstructing the permittivity-contrast function of an object from the diffracted field is one of the most celebrated inverse problems. It is encountered in research areas such as optical digital tomographic microscopy \cite{kawata1987optical,nakamura1988optical} or the characterisation of buried objects in natural environments \cite{cmielewski2007two,gurbuz2014efficient}. A widely studied physical configuration concerns an object illuminated by an incident field, produced by an emitter. The resulting diffracted field is then detected on receivers. This diffracted field represents the data of the inverse problem. They can be simulated by running the forward model (solution of the forward problem), with permittivity-contrast function as input. From a mathematical point of view, the forward model often acts as a nonlinear operator between permittivity-contrast function and data. Due to this nonlinear aspect data cannot be inverted directly. Moreover, this inverse problem is known to be ill-posed in the sense of Hadamard \cite{hadamard1902}. To circumvent this difficulty, {\it a priori} informations are introduced by different regularization procedures \cite{tikhonov1977}. The method we are extending in this paper does not address the ill-posed aspect of inverse problems, but contributes to the iterative inverse methods by introducing a new link between a variation of the permittivity-contrast function and the corresponding variation on the data. It is expressed as a functional expansion performed for any order $(n)$.\\

This expansion is reduced to its first order approximation in local optimisation methods. This is the case for the gradient method \cite{kleinman1992modified,van1997contrast} which is based on the calculus of the G\^ateaux derivative, or the Newton-Kantorovich method \cite{kantorovich1948newton} which involves the calculus of the Fr\'echet derivative. The latter can be obtained directly by differentiating the forward model \cite{roger1981newton,roger1980inverse}, or indirectly by applying the adjoint method \cite{cea1986conception}. In this last approach, the Fr\'echet derivative is constructed by solving the forward problem and its associated adjoint problem. In the latter, one considers the same object subject to a fictitious incident field, obtained by simulating a backpropagation from the receivers of the difference between the reference data and data simulated at each step of the iterative inversion process.\\

The concept of an adjoint problem has been introduced to solve inverse problems in several research areas. In electromagnetic imaging, it has been used to reconstruct the geometric shape of a cylindrical object \cite{litman1998reconstruction}. This concept has also been introduced for the purposes of acoustic imaging \cite{tarantola1984}, and subsequently generalized within the framework of elastic wave theory \cite{tarantola1987,tarantola1988}. In global seismology, inversion schemes to characterize the earth structure also involve the resolution of an adjoint problem \cite{tromp2008,monteiller2015}.\\ 

A similar approach has been introduced in electromagnetism for a configuration with emitters and receivers positioned in a farfield region, and where a harmonic time dependence is assumed \cite{roger1982}. Maxwell's equations are first introduced in the sense of distributions. Then, a calculus involving the reciprocity theorem leads to elegant forms of the Fr\'echet derivative, performed, respectively, with respect to the permittivity-contrast function and to the geometrical shape of the studied object. Such derivatives are built by solving the forward problem, and the associated reciprocal problems obtained by switching the emitter with each receiver. This principle has been applied in different inversion schemes for applications in detection of burried objects \cite{arhab2016}, optical profilometry \cite{arhab2011,arhab2012} and reconstruction of the geometric shape of three dimensional objects in microwave regime \cite{el2009adjoint}.\\

The current work aims to extend this approach to the calculus of the functional derivative for any order. The case of a three-dimensional object illuminated by an incident field in harmonic regime, is considered, where the emitter and receivers are positioned in the nearfield region \cite{yaghjian1986overview}. The paper has the following outline. First, the theoretical approach involving the reciprocity theorem \cite{roger1982} is used in section~\ref{sec:Theoretical Approach} to calculate the functional derivative of the first order, which gives the expression of the Fr\'echet derivative with respect to the permittivity-contrast function. Then the main result of this work is to show that, it is possible to apply the same approach on the remainder of the first order functional limited expansion, which gives rise to the second order functional derivative and its associated remainder. By repeating an identical calculus on this new remainder and the following ones, it becomes possible to reach the functional derivative for any order $(n)$, and then to perform a functional expansion of the data with respect to the permittivity-contrast function. The convergence of such an expansion is studied from a numerical point of view in section~\ref{sec:Application to a Forward Model}, where a two-dimensional configuration of a cylindrical object illuminated by an incident field under a ${\rm TE}$ polarisation (electric component parallel to the invariance axis), is considered. In this case, data are modelled by an operator built in the framework of the volume integral formulation. The conclusion of this work is summarized in the section~\ref{sec:Conclusion}.

\section{Theoretical Approach}
\label{sec:Theoretical Approach}
\subsection{Maxwell's Equations}
\label{subsec:Maxwell's Equations}
We consider an object of finite volume ${\mt V}$, delimited by its surface ${\mt S}$. We assume that this object is made of a dielectric (i.e. non magnetic), linear and isotropic medium. We treat the general case of an object with an inhomogeneous permittivity, described by a scalar function that depends on the position vector ${\mb r}=\sum\limits_{_{j=1}}^{_{3}} x_{_j}{\mb x}_{_j}$ in Cartesian coordinates. We introduce the complex notation, and we work under the assumption of a harmonic time dependence of the form $\propto\,{\rm e}^{{-\rm i}\omega t}$, where ${\rm i}^{^2}=-1$. For example, the source current that generates the incident field is written : ${\mb J}({\mb r},t)=\Re\{\hat{\mb J}({\mb r})\,{\rm e}^{{-\rm i}\omega t}\}$, with $\Re\{\}$ designating the real part. For the sake of clarity, from now on, the factor ${\rm e}^{{-\rm i}\omega t}$ is systematically omitted and the complex vector amplitude $\hat{\mb J}({\mb r})$ is denoted ${\mb J}({\mb r})$. We do the same for the electric field ${\mb E}$ and the magnetic field ${\mb H}$. Finally, we write Maxwell's equations in the sense of distributions. To do this, we rely on the reference \cite{appel2007} which deals with the practical aspect of distributions (For a more thorough discussion the interested reader may wish to consult the reference \cite{rudin1991}). In this way, we get :

\begin{equation}\label{Maxwell}
\left\lbrace\begin{array}{l}           
\nabla\,\times\,{\mb E}\,=\,{\rm i}\,\omega\,\mu_{_0}\,{\mb H} \\
\\
\nabla\,\times\,{\mb H}\,=\,{-\rm i}\,\omega\,\varepsilon\,{\mb E}\,+\,\delta_{{\mb r}_{\rm o}} \, {\mb J} \\
\\
\forall\,\,{\mb r}\in\,$the whole space$
\end{array}\right.
\end{equation}
 
Here the term $\delta_{{\mb r}_{\rm o}}\equiv\delta({\mb r}-{\mb r}_{\rm o})$ is the point-like Dirac distribution. We still have $\varepsilon\,=\,\varepsilon_{_0}\,\varepsilon_{\rm r}$, with $\varepsilon_{\rm r}({\mb r})$ for ${\mb r}\in\mathcal{V}$, denoting the inhomogeneous relative permittivity and where $(\varepsilon_{_0},\mu_{_0})$ are respectively permittivity and permeability of the vacuum.

\subsection{Calculus Method}
\label{subsec:Calculus Method}
\begin{figure}[htpb]
\centering
\begin{psfrags}
\psfrag{1}[][]{\scalebox{0.7}{${\rm O}$}}
\psfrag{2}[][]{\scalebox{0.7}{${\mb r}_l$}}
\psfrag{3}[][]{\scalebox{0.7}{${\mb r}_m$}}
\psfrag{4}[][]{\scalebox{0.7}{$\varepsilon_{\rm a}({\mb r})$}}
\psfrag{5}[][]{\scalebox{0.7}{$\varepsilon_{\rm b}({\mb r})$}}
\psfrag{6}[][]{\scalebox{0.7}{($\mathcal{V}$)}}
\psfrag{7}[][]{\scalebox{0.7}{($\mathcal{S}$)}}
\psfrag{8}[][]{\scalebox{0.7}{${\mb J}_l$}}
\psfrag{9}[][]{\scalebox{0.7}{${\mb E}_{{\rm a},l}({\mb r}_m)$}}
\psfrag{10}[][]{\scalebox{0.7}{${\mb E}_{{\rm b},m}({\mb r}_l)$}}
\psfrag{11}[][]{\scalebox{0.7}{${\mb J}_m$}}
\psfrag{12}[][]{\scalebox{0.7}{${\mb E}_{{\rm b},l}({\mb r}_m)$}}
\psfrag{13}[][]{\scalebox{0.7}{${\mb E}_{{\rm a},m}({\mb r}_l)$}}
\psfrag{14}[][]{\scalebox{0.7}{${\mb r}$}}
\psfrag{(a)}[][]{\scalebox{1.0}{(${\rm a}_{_l}$)}}
\psfrag{(b)}[][]{\scalebox{1.0}{(${\rm b}_{_m}$)}}
\psfrag{(b')}[][]{\scalebox{1.0}{(${\rm b}_{_l}$)}}
\psfrag{(a')}[][]{\scalebox{1.0}{(${\rm a}_{_m}$)}}
\includegraphics[height=10cm,width=15cm]{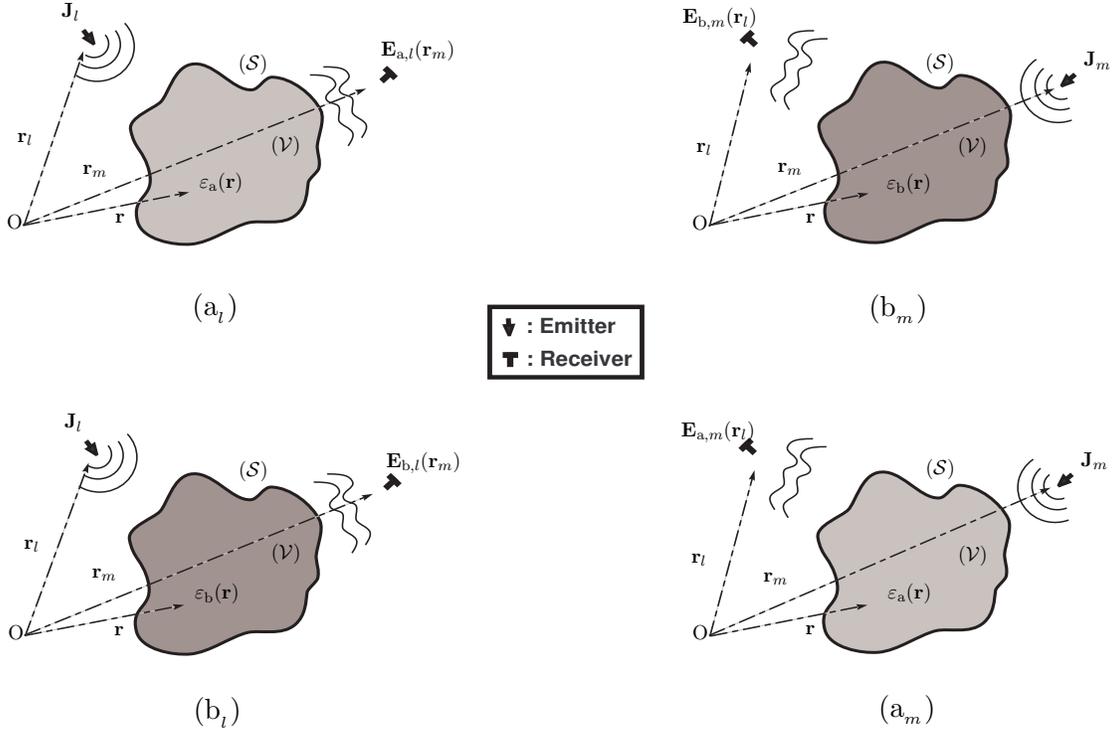} 
\caption{(${\rm a}_{_l}$), (${\rm b}_{_m}$), (${\rm b}_{_l}$) and (${\rm a}_{_m}$): The four independent physical configurations of the problem. Each one of them is fixed by the permittivity-function $\varepsilon_{\rm v}$ (${\rm v}={\rm a},{\rm b}$), and the source current ${\mb J}_{w}$ ($w=l,m$). ${\mb E}_{{\rm v},w}$ denotes the corresponding total electric field. ($\mathcal{V}$) and ($\mathcal{S}$) are respectively the volume and the surface of the studied object.
\label{configurations}}
\end{psfrags}
\end{figure}

Let us now introduce the four physical configurations (${\rm a}_{_l}$), (${\rm b}_{_m}$), (${\rm b}_{_l}$) and (${\rm a}_{_m}$) illustrated in figure Fig~(\ref{configurations}). Note that they are physically independent of each other. Each one of them is described by the permittivity-function $\varepsilon_{\rm v}({\mb r})$ (${\rm v}={\rm a},{\rm b}$ and ${\mb r}\in\mathcal{V}$), by the source current ${\mb J}_{w}$ ($w=l,m$) located at ${\mb r}_w$ and by the total electric field ${\mb E}_{{\rm v},w}$. The latter is given by the sum of the incident field generated by the source current ${\mb J}_{w}$ and the field diffracted by the studied object, namely ${\mb E}_{{\rm v},w}={\mb E}^{\rm i}_w+{\mb E}^{\rm d}_{{\rm v},w}$. Then, the system of coupled equations Eqs~(\ref{Maxwell}), can be rewritten for the generic configuration (${\rm v}_{_w}$), in the following form:

\begin{equation}\label{Maxwellv}
({\rm v}_{_w}),\,\left\lbrace\begin{array}{l}           
\nabla\,\times\,{\mb E}_{{\rm v},w}\,=\,{\rm i}\,\omega\,\mu_{_0}\,{\mb H}_{{\rm v},w} \\
\\
\nabla\,\times\,{\mb H}_{{\rm v},w}\,=\,{-\rm i}\,\omega\,\varepsilon_{\rm v}\,{\mb E}_{{\rm v},w}\,+\,\delta_{{\mb r}_w} \, {\mb J}_w \\
\\
\forall\,\,{\mb r}\in\,\text{the whole space, with}\,{\rm v}={\rm a},{\rm b}\,\text{and}\,w=l,m
\end{array}\right.
\end{equation}

From elementary vector calculus, we obtain for the electromagnetic fields of configurations (${\rm a}_{_l}$) and (${\rm b}_{_m}$) the following relation:

\begin{equation}\label{Vectorielle}
\begin{array}{l}
\nabla\,\cdot\,[{\mb H}_{{\rm b},m}\,\times\,{\mb E}_{{\rm a},l}\,-\,{\mb H}_{{\rm a},l}\,\times\,{\mb E}_{{\rm b},m}]\,=\,[{\mb E}_{{\rm a},l}\,\cdot\,(\nabla\,\times\,{\mb H}_{{\rm b},m})\,-\,{\mb H}_{{\rm b},m}\,\cdot\,(\nabla\,\times\,{\mb E}_{{\rm a},l})]\,-\, \\
\,\,\,\,\,\,\,\,\,\,\,\,\,\,\,\,\,\,\,\,\,\,\,\,\,\,\,\,\,\,\,\,\,\,\,\,\,\,\,\,\,\,\,\,\,\,\,\,\,\,\,\,\,\,[{\mb E}_{{\rm b},m}\,\cdot\,(\nabla\,\times\,{\mb H}_{{\rm a},l})\,-\,{\mb H}_{{\rm a},l}\,\cdot\,(\nabla\,\times\,{\mb E}_{{\rm b},m})]
\end{array}
\end{equation} 

By replacing the curl of the electric and magnetic fields, by their respective expressions given in Eq~(\ref{Maxwellv}), we obtain:

\begin{equation}\label{Distribution}
\nabla\,\cdot\,[{\mb H}_{{\rm b},m}\,\times\,{\mb E}_{{\rm a},l}\,-\,{\mb H}_{{\rm a},l}\,\times\,{\mb E}_{{\rm b},m}]\,=\,{-\rm i}\,\omega\,(\varepsilon_{\rm b}\,-\,\varepsilon_{\rm a})\,{\mb E}_{{\rm a},l}\,\cdot\,{\mb E}_{{\rm b},m}\,+\,\delta_{{\mb r}_m}\,{\mb E}_{{\rm a},l}\,\cdot\,{\mb J}_m\,-\,\delta_{{\mb r}_l}\,{\mb E}_{{\rm b},m}\,\cdot\,{\mb J}_l
\end{equation} 
 
Each term of the above equation is a distribution, that we can apply on a test function. After some lengthy calculations (see appendix~\ref{appendixa}), this equation is rewritten in the sense of functions, in the form:

\begin{equation}\label{Fonction1}
\begin{array}{l}
\int\limits_{\rm C}\,[{\mb H}_{{\rm b},m}\,\times\,{\mb E}_{{\rm a},l}\,-\,{\mb H}_{{\rm a},l}\,\times\,{\mb E}_{{\rm b},m}]\,\cdot\,{\mb e}_{_{\rho}}\,\,{\rm d}{\rm C}\,=\,{-\rm i}\,\omega\int\limits_{\mathcal{V}}\,(\varepsilon_{\rm b}\,-\,\varepsilon_{\rm a})\,{\mb E}_{{\rm a},l}\,\cdot\,{\mb E}_{{\rm b},m}\,\,{\rm d}\mathcal{V}\,+\,{\mb E}_{{\rm a},l}({\mb r}_m)\,\cdot\,{\mb J}_m\,-\, \\
\,\,\,\,\,\,\,\,\,\,\,\,\,\,\,\,\,\,\,\,\,\,\,\,\,\,\,\,\,\,\,\,\,\,\,\,\,\,\,\,\,\,\,\,\,\,\,\,\,\,\,\,\,\,\,\,\,\,\,\,\,\,\,\,\,\,\,\,\,\,\,\,\,\,\,\,\,\,\,\,\,\,\,\,\,\,\,\,\,\,\,\,\,\,\,\,\,\,\,\,\,\,\,\,\,\,\,\,\,\,\,\,\,\,\,{\mb E}_{{\rm b},m}({\mb r}_l)\,\cdot\,{\mb J}_l
\end{array}
\end{equation}

The left-hand side of the above equation in nil (see appendix~\ref{appendixb}), so we have:

\begin{equation}\label{Fonction2}
{\mb E}_{{\rm b},m}({\mb r}_l)\,\cdot\,{\mb J}_l\,-\,{\mb E}_{{\rm a},l}({\mb r}_m)\,\cdot\,{\mb J}_m\,=\,{-\rm i}\,\omega\int\limits_{\mathcal{V}}\,(\varepsilon_{\rm b}\,-\,\varepsilon_{\rm a})\,{\mb E}_{{\rm a},l}\,\cdot\,{\mb E}_{{\rm b},m}\,\,{\rm d}\mathcal{V}
\end{equation}

In the absence of the scatterers of permittivities $\varepsilon_{\rm a}$ and $\varepsilon_{\rm b}$, this equation is rewritten only with the incident fields:

\begin{equation}\label{Fonction3}
{\mb E}^{\rm i}_m({\mb r}_l)\,\cdot\,{\mb J}_l\,-\,{\mb E}^{\rm i}_l({\mb r}_m)\,\cdot\,{\mb J}_m\,=\,0
\end{equation} 

By subtracting (\ref{Fonction3}) from (\ref{Fonction2}), we get:

\begin{equation}\label{Fonction4}
{\mb E}^{\rm d}_{{\rm b},m}({\mb r}_l)\,\cdot\,{\mb J}_l\,-\,{\mb E}^{\rm d}_{{\rm a},l}({\mb r}_m)\,\cdot\,{\mb J}_m\,=\,{-\rm i}\,\omega\int\limits_{\mathcal{V}}\,(\varepsilon_{\rm b}\,-\,\varepsilon_{\rm a})\,{\mb E}_{{\rm a},l}\,\cdot\,{\mb E}_{{\rm b},m}\,\,{\rm d}\mathcal{V}
\end{equation}

By replacing in the above equation the configuration (${\rm a}_{_l}$) by the configuration (${\rm b}_{_l}$), we get the following equation which is nothing but the reciprocity theorem between the configurations (${\rm b}_{_l}$) and (${\rm b}_{_m}$):

\begin{equation}\label{Reciprocite}
{\mb E}^{\rm d}_{{\rm b},m}({\mb r}_l)\,\cdot\,{\mb J}_l\,-\,{\mb E}^{\rm d}_{{\rm b},l}({\mb r}_m)\,\cdot\,{\mb J}_m\,=\,0
\end{equation}

Once this equality is introduced in equation Eq~(\ref{Fonction4}), the latter can be rewritten in the form:

\begin{equation}\label{Fonction5}
[{\mb E}^{\rm d}_{{\rm b},l}({\mb r}_m)\,-\,{\mb E}^{\rm d}_{{\rm a},l}({\mb r}_m)]\,\cdot\,{\mb J}_m\,=\,{-\rm i}\,\omega\int\limits_{\mathcal{V}}\,(\varepsilon_{\rm b}\,-\,\varepsilon_{\rm a})\,{\mb E}_{{\rm a},l}\,\cdot\,{\mb E}_{{\rm b},m}\,\,{\rm d}\mathcal{V}
\end{equation}

To derive the variational form of expression Eq~(\ref{Fonction5}), we add the assumption that $\varepsilon_{\rm b}\,=\,\varepsilon_{\rm a}\,+\,\delta\varepsilon$. This implies, for the fields of the four studied configurations, the following relations: ${\mb E}^{\rm d}_{{\rm b},l}({\mb r}_m)\,=\,{\mb E}^{\rm d}_{{\rm a},l}({\mb r}_m)\,+\,\delta{\mb E}^{\rm d}_{{\rm a},l}({\mb r}_m)$ and ${\mb E}_{{\rm b},m}({\mb r})\,=\,{\mb E}_{{\rm a},m}({\mb r})\,+\,\delta{\mb E}_{{\rm a},m}({\mb r})$. It follows that the expression Eq~(\ref{Fonction5}) can be rewritten as:

\begin{equation}
\delta{\mb E}^{\rm d}_{{\rm a},l}({\mb r}_m)\,\cdot\,{\mb J}_m\,=\,{-\rm i}\omega\int\limits_{\mathcal{V}}\,\delta\varepsilon({\mb r})\,\,{\mb E}_{{\rm a},l}({\mb r})\,\cdot\,{\mb E}_{{\rm a},m}({\mb r})\,\,{\rm d}\mathcal{V}\,{-\rm i}\omega\int\limits_{\mathcal{V}}\,\delta\varepsilon({\mb r})\,\,{\mb E}_{{\rm a},l}({\mb r})\,\cdot\,\delta{\mb E}_{{\rm a},m}({\mb r})\,\,{\rm d}\mathcal{V}
\end{equation}
 
The electric fields of the above expression are all calculated for the same permittivity $\varepsilon_{\rm a}$. Thus, for the sake of clarity, we omit the index $(a)$. Moreover, the introduction into the same expression of the relations: $\varepsilon\,=\,\varepsilon_{_0}\,\varepsilon_{\rm r}$ and $\chi\,=\,\varepsilon_{\rm r}\,-\,1$, leads us to rewrite it according to the permittivity-contrast function $\chi$: 

\begin{equation}\label{EgaliteDeBase}
\left\lbrace\begin{array}{l}     
\delta{\mb E}^{\rm d}_l({\mb r}_m)\,\cdot\,{\mb J}_m\,=\,{-\rm i}\omega\varepsilon_{_0}\int\limits_{\mathcal{V}}\,\delta\chi({\mb r})\,\,{\mb E}_l({\mb r})\,\cdot\,{\mb E}_m({\mb r})\,\,{\rm d}\mathcal{V}\,{-\rm i}\omega\varepsilon_{_0}\int\limits_{\mathcal{V}}\,\delta\chi({\mb r})\,\,{\mb E}_l({\mb r})\,\cdot\,\delta{\mb E}_m({\mb r})\,\,{\rm d}\mathcal{V}\\
\Leftrightarrow\,\,\,\,\,\,\delta\mathcal{D}_{lm}\,=\,\mathcal{F}_{_{lm}}^{^{(1)}}(\delta\chi)\,+\,{\rm o}_{_{lm}}(||\delta\chi||) ,\,\,\,\,\,\,\,\,\,\,\,\,\,\,\,\,\,\,\,\,\,\,\,\,\,\text{(in a compact form)}
\end{array}\right. 
\end{equation}
According to the left-hand side of the above expression, data element $\delta\mathcal{D}_{lm}$ is given by the dot product between the variation of the diffracted field $\delta{\mb E}^{\rm d}_l$ at ${\mb r}_m$ (this field is diffracted by the object after interacting with the incident field emitted by the source current ${\mb J}_l$ located at ${\mb r}_l$) and the source current ${\mb J}_m$ that would be positioned at ${\mb r}_m$ (This source current radiates an incident field that interacts with the object to give the total electric field ${\mb E}_m({\mb r}),\,\text{for}\,{\mb r}\in\mathcal{V}$). The above calculus was first proposed in farfield configuration \cite{roger1982}, in this context the remainder ${\rm o}_{_{lm}}(||\delta\chi||)$ has been neglected. In this work, this calculus is adapted to localized emitters and receivers. Now, from the right-hand side of the above expression, and by taking into account the remainder ${\rm o}_{_{lm}}(||\delta\chi||)$, it is now possible to extract the functional derivatives up to any order $(n)$. 

\subsection{Functional derivatives of orders $(1)$, $(2)$ and $(3)$}
\label{subsec:Functional derivatives of orders $(1)$, $(2)$ and $(3)$}
\begin{itemize}
\item[$\bullet$] \textbf{First order functional derivative (Fr\'echet derivative)} This derivative which appears directly in expression Eq~(\ref{EgaliteDeBase}), is obtained by neglecting the remainder ${\rm o}_{_{lm}}(||\delta\chi||)$ that contains all the derivatives of higher orders:

\begin{equation}\label{Fonctionnelle1}
\left\lbrace\begin{array}{l}
\mathcal{F}_{_{lm}}^{^{(1)}}(\delta\chi)\,=\,{-\rm i}\omega\varepsilon_{_0}\int\limits_{\mathcal{V}}\,\delta\chi({\mb r})\,\,{\mb E}_l({\mb r})\,\cdot\,{\mb E}_m({\mb r})\,\,{\rm d}\mathcal{V} \\
\text{Where $\mathcal{F}_{_{lm}}^{^{(1)}}$ denotes the first order functional derivative}
\end{array}\right.
\end{equation}

\item[$\bullet$] \textbf{Second order functional derivative} This second derivative is obtained by writing: ${\rm o}_{_{lm}}(||\delta\chi||)\,=\,\mathcal{F}_{_{lm}}^{^{(2)}}(\delta\chi,\delta\chi)\,+\,{\rm o}_{_{lm}}(||\delta\chi||^{^{2}})$. To deduce the explicit form of $\mathcal{F}_{_{lm}}^{^{(2)}}$, let us start with the expression of ${\rm o}_{_{lm}}(||\delta\chi||)$:

\begin{equation}\label{Fonctionnelle2-1}
{\rm o}_{_{lm}}(||\delta\chi||)\,=\,{-\rm i}\omega\varepsilon_{_0}\int\limits_{\mathcal{V}}\,\delta\chi({\mb r})\,\,{\mb E}_l({\mb r})\,\cdot\,\delta{\mb E}_m({\mb r})\,\,{\rm d}\mathcal{V}\,=\,\int\limits_{\mathcal{V}}\,\delta{\mb E}_m({\mb r})\,\cdot\,[{-\rm i}\omega\varepsilon_{_0}\,\delta\chi({\mb r})\,{\mb E}_l({\mb r})\,\,{\rm d}\mathcal{V}]
\end{equation}

The novelty of this work (to the best of the authors' knowledge) is to set $[{-\rm i}\omega\varepsilon_{_0}\,\delta\chi({\mb r})\,{\mb E}_l({\mb r})\,\,{\rm d}\mathcal{V}]\,=\,{\mb J}_l({\mb r})$. Indeed, given its physical dimension, this term can be assimilated to a source current. However, to make the calculus more convenient, we introduce the reduced source current ${\bar{\mb J}}_l({\mb r})$, such that: ${\mb J}_l({\mb r})\,=\,{\bar{\mb J}}_l({\mb r})\,{\rm d}\mathcal{V}$. This allows us to rewrite the expression Eq~(\ref{Fonctionnelle2-1}) as: 

\begin{equation}\label{Fonctionnelle2-2}
{\rm o}_{_{lm}}(||\delta\chi||)\,=\,{-\rm i}\omega\varepsilon_{_0}\int\limits_{\mathcal{V}}\,\delta\chi({\mb r})\,\,{\mb E}_l({\mb r})\,\cdot\,\delta{\mb E}_m({\mb r})\,\,{\rm d}\mathcal{V}\,=\,\int\limits_{\mathcal{V}}\,\delta{\mb E}_m({\mb r})\,\cdot\,{\bar{\mb J}}_l({\mb r})\,\,{\rm d}\mathcal{V}
\end{equation}

Based on equation Eq~(\ref{EgaliteDeBase}), this integral can be transformed into:

\begin{equation}\label{Fonctionnelle2-3}
\begin{array}{l}
\int\limits_{\mathcal{V}}\,\delta{\mb E}_m({\mb r})\,\cdot\,{\bar{\mb J}}_l({\mb r})\,\,{\rm d}\mathcal{V}\,=\,\int\limits_{\mathcal{V}}\,\{{-\rm i}\omega\varepsilon_{_0}\int\limits_{\mathcal{V}}\,\delta\chi({\mb r'})\,\,{\mb E}_m({\mb r'})\,\cdot\,\bar{{\mb E}}_l({\mb r},\,{\mb r'})\,\,{\rm d}\mathcal{V'}\, \\
\,\,\,\,\,\,\,\,\,\,\,\,\,\,\,\,\,\,\,\,\,\,\,\,\,\,\,\,\,\,\,\,\,\,\,\,\,\,\,\,\,\,\,\,\,\,\,\,\,\,\,\,\,\,\,\,\,\,\,\,\,\,\,\,\,\,\,\,\,\,\,\,\,\,\,\,\,\,\,\,\,\,\,\,\,\,\,\,{-\rm i}\omega\varepsilon_{_0}\int\limits_{\mathcal{V}}\,\delta\chi({\mb r'})\,\,{\mb E}_m({\mb r'})\,\cdot\,\delta\bar{{\mb E}}_l({\mb r},\,{\mb r'})\,\,{\rm d}\mathcal{V'}\}\,\,{\rm d}\mathcal{V}
\end{array}
\end{equation}
 
In this new expression, $\bar{{\mb E}}_l({\mb r},\,{\mb r'})$ is the reduced electric field calculated at the point ${\mb r'}$. It is the result of an interaction between the studied object and a fictitious incident field, generated by the reduced source current ${\bar{\mb J}}_l({\mb r})$ located at the point ${\mb r}$. Eq~(\ref{Fonctionnelle2-3}) can also be rewritten in the form:

\begin{equation}\label{Fonctionnelle2-4}
\begin{array}{l}
\int\limits_{\mathcal{V}}\,\delta{\mb E}_m({\mb r})\,\cdot\,{\bar{\mb J}}_l({\mb r})\,\,{\rm d}\mathcal{V}\,=\,{-\rm i}\omega\varepsilon_{_0}\int\limits_{\mathcal{V}}\,\delta\chi({\mb r'})\,\,{\mb E}_m({\mb r'})\,\cdot\,\{\int\limits_{\mathcal{V}}\bar{{\mb E}}_l({\mb r},\,{\mb r'})\,{\rm d}\mathcal{V}\}\,\,{\rm d}\mathcal{V'}\, \\
\,\,\,\,\,\,\,\,\,\,\,\,\,\,\,\,\,\,\,\,\,\,\,\,\,\,\,\,\,\,\,\,\,\,\,\,\,\,\,\,\,\,\,\,\,\,\,\,\,\,\,\,\,\,\,\,\,\,\,\,\,\,\,\,\,\,\,\,\,\,\,\,\,\,\,\,\,\,\,\,\,\,\,\,\,\,\,\,{-\rm i}\omega\varepsilon_{_0}\int\limits_{\mathcal{V}}\,\delta\chi({\mb r'})\,\,{\mb E}_m({\mb r'})\,\cdot\,\{\int\limits_{\mathcal{V}}\delta\bar{{\mb E}}_l({\mb r},\,{\mb r'})\,{\rm d}\mathcal{V}\}\,\,{\rm d}\mathcal{V'}
\end{array}
\end{equation}

In the above expression, the integral $\{\int\limits_{\mathcal{V}}\bar{{\mb E}}_l({\mb r},\,{\mb r'})\,{\rm d}\mathcal{V}\}={\mb E}_{l,{\mb r}^{*}_{\scalebox{0.4}{(1)}}}({\mb r'})$ is the resulting electric field, due to an interaction between the studied object and all the fictitious incident fields generated by the set of source currents ${\mb J}_l({\mb r})$, where ${\mb r}$ spans the domain $\mathcal{V}$. Here, the index ${\mb r}^{*}_{\scalebox{0.4}{(1)}}$ has two significations: the symbol $*$ means that we integrate in the domain with respect to the position ${\mb r}$, in order to take precisely into account the contributions of all the source currents in the calculation of the total field. The index $(1)$ means that the variation of the permittivity-contrast function is implicitly embedded once in the total electric field ${\mb E}_{l,{\mb r}^{*}_{\scalebox{0.4}{(1)}}}$. Then, Eq~(\ref{Fonctionnelle2-4}) is rewritten in the form:

\begin{equation}\label{Fonctionnelle2-5}
\begin{array}{l}
\int\limits_{\mathcal{V}}\,\delta{\mb E}_m({\mb r})\,\cdot\,{\bar{\mb J}}_l({\mb r})\,\,{\rm d}\mathcal{V}\,=\,{-\rm i}\omega\varepsilon_{_0}\int\limits_{\mathcal{V}}\,\delta\chi({\mb r'})\,\,{\mb E}_m({\mb r'})\,\cdot\,{\mb E}_{l,{\mb r}^{*}_{\scalebox{0.4}{(1)}}}({\mb r'})\,\,{\rm d}\mathcal{V'}\, \\
\,\,\,\,\,\,\,\,\,\,\,\,\,\,\,\,\,\,\,\,\,\,\,\,\,\,\,\,\,\,\,\,\,\,\,\,\,\,\,\,\,\,\,\,\,\,\,\,\,\,\,\,\,\,\,\,\,\,\,\,\,\,\,\,\,\,\,\,\,\,\,\,\,\,\,\,\,\,\,\,\,\,\,\,\,\,\,\,{-\rm i}\omega\varepsilon_{_0}\int\limits_{\mathcal{V}}\,\delta\chi({\mb r'})\,\,{\mb E}_m({\mb r'})\,\cdot\,\delta{\mb E}_{l,{\mb r}^{*}_{\scalebox{0.4}{(1)}}}({\mb r'})\,\,{\rm d}\mathcal{V'}
\end{array}
\end{equation}

To simplify the notation, let's replace ${\mb r'}$ by ${\mb r}$ in the right hand side of the above equation. Eqs~(\ref{Fonctionnelle2-2},\ref{Fonctionnelle2-5}) lead to the following result:

\begin{equation}\label{PetitO1}
\left\lbrace\begin{array}{l}     
{-\rm i}\omega\varepsilon_{_0}\int\limits_{\mathcal{V}}\,\delta\chi({\mb r})\,\,{\mb E}_l({\mb r})\,\cdot\,\delta{\mb E}_m({\mb r})\,\,{\rm d}\mathcal{V}\,=\,{-\rm i}\omega\varepsilon_{_0}\int\limits_{\mathcal{V}}\,\delta\chi({\mb r})\,\,{\mb E}_m({\mb r})\,\cdot\,{\mb E}_{l,{\mb r}^{*}_{\scalebox{0.4}{(1)}}}({\mb r})\,\,{\rm d}\mathcal{V}\,\\
\,\,\,\,\,\,\,\,\,\,\,\,\,\,\,\,\,\,\,\,\,\,\,\,\,\,\,\,\,\,\,\,\,\,\,\,\,\,\,\,\,\,\,\,\,\,\,\,\,\,\,\,\,\,\,\,\,\,\,\,\,\,\,\,\,\,\,\,\,\,\,\,\,\,\,\,\,\,\,\,\,\,\,\,\,\,\,\,\,\,\,\,\,\,\,\,\,\,\,\,\,\,\,\,\,\,\,\,\,\,\,\,\,\,\,\,\,\,\,\,\,\,\,\,\,\,\,\,\,\,\,\,\,\,\,\,\,\,\,\,\,\,{-\rm i}\omega\varepsilon_{_0}\int\limits_{\mathcal{V}}\,\delta\chi({\mb r})\,\,{\mb E}_m({\mb r})\,\cdot\,\delta{\mb E}_{l,{\mb r}^{*}_{\scalebox{0.4}{(1)}}}({\mb r})\,\,{\rm d}\mathcal{V}\\
\Leftrightarrow\,\,\,\,\,\,{\rm o}_{_{lm}}(||\delta\chi||)\,=\,\mathcal{F}_{_{lm}}^{^{(2)}}(\delta\chi,\delta\chi)\,+\,{\rm o}_{_{lm}}(||\delta\chi||^{^{2}}) ,\,\,\,\,\,\,\,\,\,\,\,\,\,\,\,\,\,\,\,\,\,\,\,\,\,\,\,\,\,\,\,\,\,\,\,\,\,\,\,\,\,\,\text{(in a compact form)}
\end{array}\right.
\end{equation}

With:

\begin{equation}\label{Fonctionnelle2-6}
\left\lbrace\begin{array}{l}
\mathcal{F}_{_{lm}}^{^{(2)}}(\delta\chi,\delta\chi)\,=\,{-\rm i}\omega\varepsilon_{_0}\int\limits_{\mathcal{V}}\,\delta\chi({\mb r})\,\,{\mb E}_m({\mb r})\,\cdot\,{\mb E}_{l,{\mb r}^{*}_{\scalebox{0.4}{(1)}}}({\mb r})\,\,{\rm d}\mathcal{V} \\
\text{Where $\mathcal{F}_{_{lm}}^{^{(2)}}$ is the second order functional derivative}
\end{array}\right.
\end{equation}

\item[$\bullet$] \textbf{Third order functional derivative} To find the expression of this derivative, we repeat the same calculus scheme, i.e by writing : ${\rm o}_{_{lm}}(||\delta\chi||^{^{2}})\,=\,\mathcal{F}_{_{lm}}^{^{(3)}}(\delta\chi,\delta\chi,\delta\chi)\,+\,{\rm o}_{_{lm}}(||\delta\chi||^{^{3}})$. So, we start from the expression of the remainder ${\rm o}_{_{lm}}(||\delta\chi||^{^{2}})$: 

\begin{equation}\label{Fonctionnelle3-1}
{\rm o}_{_{lm}}(||\delta\chi||^{^{2}})\,=\,{-\rm i}\omega\varepsilon_{_0}\int\limits_{\mathcal{V}}\,\delta\chi({\mb r})\,\,{\mb E}_m({\mb r})\,\cdot\,\delta{\mb E}_{l,{\mb r}^{*}_{\scalebox{0.4}{(1)}}}({\mb r})\,\,{\rm d}\mathcal{V}\,=\,\int\limits_{\mathcal{V}}\,\delta{\mb E}_{l,{\mb r}^{*}_{\scalebox{0.4}{(1)}}}({\mb r})\,\cdot\,[{-\rm i}\omega\varepsilon_{_0}\,\delta\chi({\mb r})\,{\mb E}_m({\mb r})\,\,{\rm d}\mathcal{V}]
\end{equation}
 
By following the same procedure one obtains:

\begin{equation}\label{PetitO2}
\left\lbrace\begin{array}{l}     
{-\rm i}\omega\varepsilon_{_0}\int\limits_{\mathcal{V}}\,\delta\chi({\mb r})\,\,{\mb E}_m({\mb r})\,\cdot\,\delta{\mb E}_{l,{\mb r}^{*}_{\scalebox{0.4}{(1)}}}({\mb r})\,\,{\rm d}\mathcal{V}\,=\,{-\rm i}\omega\varepsilon_{_0}\int\limits_{\mathcal{V}}\,\delta\chi({\mb r})\,\,{\mb E}_{l,{\mb r}^{*}_{\scalebox{0.4}{(1)}}}({\mb r})\,\cdot\,{\mb E}_{m,{\mb r}^{*}_{\scalebox{0.4}{(1)}}}({\mb r})\,\,{\rm d}\mathcal{V}\,\\
\,\,\,\,\,\,\,\,\,\,\,\,\,\,\,\,\,\,\,\,\,\,\,\,\,\,\,\,\,\,\,\,\,\,\,\,\,\,\,\,\,\,\,\,\,\,\,\,\,\,\,\,\,\,\,\,\,\,\,\,\,\,\,\,\,\,\,\,\,\,\,\,\,\,\,\,\,\,\,\,\,\,\,\,\,\,\,\,\,\,\,\,\,\,\,\,\,\,\,\,\,\,\,\,\,\,\,\,\,\,\,\,\,\,\,\,\,\,\,\,\,\,\,\,\,\,\,\,\,\,\,\,\,\,\,\,\,\,\,\,\,\,{-\rm i}\omega\varepsilon_{_0}\int\limits_{\mathcal{V}}\,\delta\chi({\mb r})\,\,{\mb E}_{l,{\mb r}^{*}_{\scalebox{0.4}{(1)}}}({\mb r})\,\cdot\,\delta{\mb E}_{m,{\mb r}^{*}_{\scalebox{0.4}{(1)}}}({\mb r})\,\,{\rm d}\mathcal{V}\\
\Leftrightarrow\,\,\,\,\,\,{\rm o}_{_{lm}}(||\delta\chi||^{^{2}})\,=\,\mathcal{F}_{_{lm}}^{^{(3)}}(\delta\chi,\delta\chi,\delta\chi)\,+\,{\rm o}_{_{lm}}(||\delta\chi||^{^{3}}) ,\,\,\,\,\,\,\,\,\,\,\,\,\,\,\,\,\,\,\,\,\,\,\,\,\,\,\,\,\,\,\,\,\,\,\,\,\,\,\,\,\,\,\text{(in a compact form)}
\end{array}\right. 
\end{equation}

With:

\begin{equation}\label{Fonctionnelle3-2}
\left\lbrace\begin{array}{l}
\mathcal{F}_{_{lm}}^{^{(3)}}(\delta\chi,\delta\chi,\delta\chi)\,=\,{-\rm i}\omega\varepsilon_{_0}\int\limits_{\mathcal{V}}\,\delta\chi({\mb r})\,\,{\mb E}_{l,{\mb r}^{*}_{\scalebox{0.4}{(1)}}}({\mb r})\,\cdot\,{\mb E}_{m,{\mb r}^{*}_{\scalebox{0.4}{(1)}}}({\mb r})\,\,{\rm d}\mathcal{V} \\
\text{Where $\mathcal{F}_{_{lm}}^{^{(3)}}$ is the third functional derivative}
\end{array}\right.
\end{equation}
\end{itemize}

\subsection{Functional derivative of order $(n)$} The previous results can be generalized to the calculus of the functional derivative for any order $n$. We treat the two cases where $n$ is even $n=2p$ or odd $n=2p+1$, with $p\geqslant 2$ a natural number. 

\medskip
\begin{itemize}
\item[$\bullet$] \textbf{Functional derivative of order $(2p)$} Its expression is given by:

\begin{equation}\label{PetitO2pmoins1}
\left\lbrace\begin{array}{l}     
{-\rm i}\omega\varepsilon_{_0}\int\limits_{\mathcal{V}}\,\delta\chi({\mb r})\,\,{\mb E}_{l,{\mb r}^{*}_{\scalebox{0.4}{(p-1)}}}({\mb r})\,\cdot\,\delta{\mb E}_{m,{\mb r}^{*}_{\scalebox{0.4}{(p-1)}}}({\mb r})\,\,{\rm d}\mathcal{V}\,=\,{-\rm i}\omega\varepsilon_{_0}\int\limits_{\mathcal{V}}\,\delta\chi({\mb r})\,\,{\mb E}_{m,{\mb r}^{*}_{\scalebox{0.4}{(p-1)}}}({\mb r})\,\cdot\,{\mb E}_{l,{\mb r}^{*}_{\scalebox{0.4}{(p)}}}({\mb r})\,\,{\rm d}\mathcal{V}\,\\
\,\,\,\,\,\,\,\,\,\,\,\,\,\,\,\,\,\,\,\,\,\,\,\,\,\,\,\,\,\,\,\,\,\,\,\,\,\,\,\,\,\,\,\,\,\,\,\,\,\,\,\,\,\,\,\,\,\,\,\,\,\,\,\,\,\,\,\,\,\,\,\,\,\,\,\,\,\,\,\,\,\,\,\,\,\,\,\,\,\,\,\,\,\,\,\,\,\,\,\,\,\,\,\,\,\,\,\,\,\,\,\,\,\,\,\,\,\,\,\,\,\,\,\,\,\,\,\,\,\,\,\,\,\,\,\,\,\,\,\,\,\,{-\rm i}\omega\varepsilon_{_0}\int\limits_{\mathcal{V}}\,\delta\chi({\mb r})\,\,{\mb E}_{m,{\mb r}^{*}_{\scalebox{0.4}{(p-1)}}}({\mb r})\,\cdot\,\delta{\mb E}_{l,{\mb r}^{*}_{\scalebox{0.4}{(p)}}}({\mb r})\,\,{\rm d}\mathcal{V}\\
\Leftrightarrow\,\,\,\,\,\,{\rm o}_{_{lm}}(||\delta\chi||^{^{2p-1}})\,=\,\mathcal{F}_{_{lm}}^{^{(2p)}}(\underbrace{\delta\chi,\,...\,,\delta\chi}_{2p\,\text{times}})\,+\,{\rm o}_{_{lm}}(||\delta\chi||^{^{2p}}) ,\,\,\,\,\,\,\,\,\,\,\,\,\,\,\,\,\,\,\,\,\,\,\,\,\,\,\,\,\,\,\,\,\,\,\,\,\,\,\,\,\,\,\text{(in a compact form)}
\end{array}\right. 
\end{equation}

With:

\begin{equation}\label{Fonctionnelle2p}
\left\lbrace\begin{array}{l}
\mathcal{F}_{_{lm}}^{^{(2p)}}(\underbrace{\delta\chi,\,...\,,\delta\chi}_{2p\,\text{times}})\,=\,{-\rm i}\omega\varepsilon_{_0}\int\limits_{\mathcal{V}}\,\delta\chi({\mb r})\,\,{\mb E}_{m,{\mb r}^{*}_{\scalebox{0.4}{(p-1)}}}({\mb r})\,\cdot\,{\mb E}_{l,{\mb r}^{*}_{\scalebox{0.4}{(p)}}}({\mb r})\,\,{\rm d}\mathcal{V} \\
\text{Where $\mathcal{F}_{_{lm}}^{^{(2p)}}$ is the functional derivative of order $(2p)$}
\end{array}\right.
\end{equation}

\item[$\bullet$] \textbf{Functional derivative of order $(2p+1)$} It is written in the form: 

\begin{equation}\label{PetitO2p}
\left\lbrace\begin{array}{l}     
{-\rm i}\omega\varepsilon_{_0}\int\limits_{\mathcal{V}}\,\delta\chi({\mb r})\,\,{\mb E}_{m,{\mb r}^{*}_{\scalebox{0.4}{(p-1)}}}({\mb r})\,\cdot\,\delta{\mb E}_{l,{\mb r}^{*}_{\scalebox{0.4}{(p)}}}({\mb r})\,\,{\rm d}\mathcal{V}\,=\,{-\rm i}\omega\varepsilon_{_0}\int\limits_{\mathcal{V}}\,\delta\chi({\mb r})\,\,{\mb E}_{l,{\mb r}^{*}_{\scalebox{0.4}{(p)}}}({\mb r})\,\cdot\,{\mb E}_{m,{\mb r}^{*}_{\scalebox{0.4}{(p)}}}({\mb r})\,\,{\rm d}\mathcal{V}\,\\
\,\,\,\,\,\,\,\,\,\,\,\,\,\,\,\,\,\,\,\,\,\,\,\,\,\,\,\,\,\,\,\,\,\,\,\,\,\,\,\,\,\,\,\,\,\,\,\,\,\,\,\,\,\,\,\,\,\,\,\,\,\,\,\,\,\,\,\,\,\,\,\,\,\,\,\,\,\,\,\,\,\,\,\,\,\,\,\,\,\,\,\,\,\,\,\,\,\,\,\,\,\,\,\,\,\,\,\,\,\,\,\,\,\,\,\,\,\,\,\,\,\,\,\,\,\,\,\,\,\,\,\,\,\,\,\,\,\,\,\,\,\,{-\rm i}\omega\varepsilon_{_0}\int\limits_{\mathcal{V}}\,\delta\chi({\mb r})\,\,{\mb E}_{l,{\mb r}^{*}_{\scalebox{0.4}{(p)}}}({\mb r})\,\cdot\,\delta{\mb E}_{m,{\mb r}^{*}_{\scalebox{0.4}{(p)}}}({\mb r})\,\,{\rm d}\mathcal{V}\\
\Leftrightarrow\,\,\,\,\,\,{\rm o}_{_{lm}}(||\delta\chi||^{^{2p}})\,=\,\mathcal{F}_{_{lm}}^{^{(2p+1)}}(\underbrace{\delta\chi,\,...\,,\delta\chi}_{2p+1\,\text{times}})\,+\,{\rm o}_{_{lm}}(||\delta\chi||^{^{2p+1}}) ,\,\,\,\,\,\,\,\,\,\,\,\,\,\,\,\,\,\,\,\,\,\,\,\,\,\,\,\,\,\,\,\,\,\,\,\,\,\,\,\,\,\,\text{(in a compact form)}
\end{array}\right. 
\end{equation}

With:

\begin{equation}\label{Fonctionnelle2pplus1}
\left\lbrace\begin{array}{l}
\mathcal{F}_{_{lm}}^{^{(2p+1)}}(\underbrace{\delta\chi,\,...\,,\delta\chi}_{2p+1\,\text{times}})\,=\,{-\rm i}\omega\varepsilon_{_0}\int\limits_{\mathcal{V}}\,\delta\chi({\mb r})\,\,{\mb E}_{l,{\mb r}^{*}_{\scalebox{0.4}{(p)}}}({\mb r})\,\cdot\,{\mb E}_{m,{\mb r}^{*}_{\scalebox{0.4}{(p)}}}({\mb r})\,\,{\rm d}\mathcal{V} \\
\text{Where $\mathcal{F}_{_{lm}}^{^{(2p+1)}}$ is the functional derivative of order $(2p+1)$}
\end{array}\right.
\end{equation}
\end{itemize}
\vspace*{0.5cm}
The total electric field ${\mb E}_{w,{\mb r}^{*}_{\scalebox{0.4}{(p)}}}({\mb r})$ (with $w=l,m$) is obtained after solving $p+1$ forward problems. This resolution is illustrated by the following diagram:

\begin{figure}[htpb]
\centering
\begin{psfrags}
\psfrag{1}[][]{\scalebox{0.8}{${\mb J}_w$}}
\psfrag{2}[][]{\scalebox{0.8}{${\mb E}_w({\mb r}),\,\,{\mb r}\in\mathcal{V}$}}
\psfrag{3}[][]{\scalebox{0.8}{$\begin{array}{l}-{\rm i}\omega\varepsilon_{\rm o}\delta\chi({\mb r}){\mb E}_w({\mb r}){\rm d}\mathcal{V}, \\\,\,\,\, {\mb r}\in\mathcal{V} \end{array}$}}
\psfrag{4}[][]{\scalebox{0.8}{${\mb E}_{w,{\mb r}^{*}_{\scalebox{0.4}{(1)}}}({\mb r}),\,\,{\mb r}\in\mathcal{V}$}}
\psfrag{5}[][]{\scalebox{0.8}{$\begin{array}{l}-{\rm i}\omega\varepsilon_{\rm o}\delta\chi({\mb r}){\mb E}_{w,{\mb r}^{*}_{\scalebox{0.4}{(1)}}}({\mb r}){\rm d}\mathcal{V}, \\\,\,\,\, {\mb r}\in\mathcal{V} \end{array}$}}
\psfrag{6}[][]{\scalebox{0.8}{${\mb E}_{w,{\mb r}^{*}_{\scalebox{0.4}{(2)}}}({\mb r}),\,\,{\mb r}\in\mathcal{V}$}}
\psfrag{7}[][]{\scalebox{0.8}{$\begin{array}{l}-{\rm i}\omega\varepsilon_{\rm o}\delta\chi({\mb r}){\mb E}_{w,{\mb r}^{*}_{\scalebox{0.4}{(p-1)}}}({\mb r}){\rm d}\mathcal{V}, \\\,\,\,\, {\mb r}\in\mathcal{V} \end{array}$}}
\psfrag{8}[][]{\scalebox{0.78}{${\mb E}_{w,{\mb r}^{*}_{\scalebox{0.4}{(p)}}}({\mb r}),\,\,{\mb r}\in\mathcal{V}$}}
\psfrag{s0}[][]{\scalebox{0.7}{\underline{\textbf{step 0}}}}
\psfrag{s1}[][]{\scalebox{0.7}{\underline{\textbf{step 1}}}}
\psfrag{s2}[][]{\scalebox{0.7}{\underline{\textbf{step 2}}}}
\psfrag{sp}[][]{\scalebox{0.7}{\underline{\textbf{step p}}}}
\includegraphics[height=5.5cm,width=18.5cm]{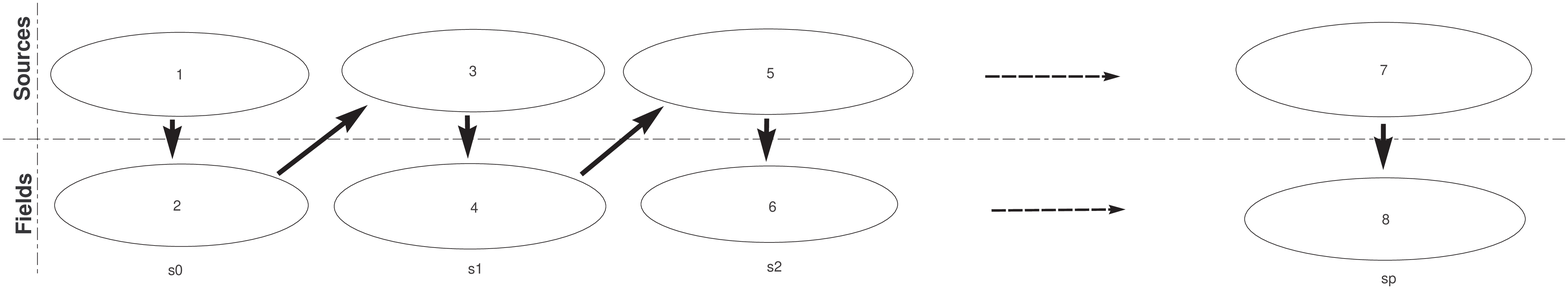} 
\caption{Resolution of $p+1$ forward problems. The source current ${\mb J}_w$ located at ${\mb r}_w$ radiates an incident field that interacts with the object, this gives rise to the total field ${\mb E}_w({\mb r})\,\text{for}\,{\mb r}\in\mathcal{V}$. Then, for each position ${\mb r}\in\mathcal{V}$, the term $-{\rm i}\omega\varepsilon_{\rm o}\delta\chi({\mb r}){\mb E}_w({\mb r}){\rm d}\mathcal{V}$ acts like a new source current that radiates a new incident field. The interaction of the object with the total incident field emitted by all the source currents $-{\rm i}\omega\varepsilon_{\rm o}\delta\chi({\mb r}){\mb E}_w({\mb r}){\rm d}\mathcal{V}$ with ${\mb r}$ spanning the domain $\mathcal{V}$, gives rise to the new total field ${\mb E}_{w,{\mb r}^{*}_{\scalebox{0.4}{(1)}}}({\mb r})\,\text{for}\,{\mb r}\in\mathcal{V}$. As shown in the above diagram, the latter is involved in a new source currents to calculate a new total field, and so on.
\label{p-plus-1-forward-problems}}
\end{psfrags}
\end{figure}

The next section as well as appendix~\ref{appendixc} describe how the volume integral formulation is used, for the numerical evaluation of the electric fields involved in the different steps of the above diagram.

\section{Application to a Forward Model}
\label{sec:Application to a Forward Model}
\subsection{Study Configuration and Incident Field}
\label{subsec:Study Configuration and Incident Field}

\begin{figure}[htpb]
\centering
\begin{psfrags}
\psfrag{1}[][]{\scalebox{0.8}{${\rm J}_{l=1}$}}
\psfrag{2}[][]{\scalebox{0.8}{${\rm E}^{\rm d}_{m=2}$}}
\psfrag{3}[][]{\scalebox{0.8}{${\rm E}^{\rm d}_{m=3}$}}
\psfrag{4}[][]{\scalebox{0.8}{${\rm E}^{\rm d}_{m={\rm M}}$}}
\psfrag{5}[][]{\scalebox{0.7}{$5\,{\rm cm}$}}
\psfrag{6}[][]{\scalebox{0.7}{$11\,{\rm cm}$}}
\psfrag{7}[][]{\scalebox{0.8}{$\begin{array}{l}(\varepsilon_{_0}\varepsilon_{\rm r},\,\mu_{_0})\\\chi=\varepsilon_{\rm r}-1\end{array}$}}
\psfrag{8}[][]{\scalebox{0.8}{${\mb x}_{_1}$}}
\psfrag{9}[][]{\scalebox{0.8}{${\mb x}_{_3}$}}
\psfrag{10}[][]{\scalebox{0.8}{${\mb x}_{_2}$}}
\psfrag{11}[][]{\scalebox{0.8}{: invariance axis}}
\psfrag{12}[][]{\scalebox{0.8}{$(\varepsilon_{_0},\,\mu_{_0})$}}
\psfrag{13}[][]{\scalebox{0.7}{Emitter}}
\psfrag{14}[][]{\scalebox{0.7}{Receiver}}
\psfrag{15}[][]{\scalebox{1.0}{$(\mathcal{V})$}}
\psfrag{16}[][]{\scalebox{1.0}{$(\Gamma)$}}
\includegraphics[height=9cm,width=12cm]{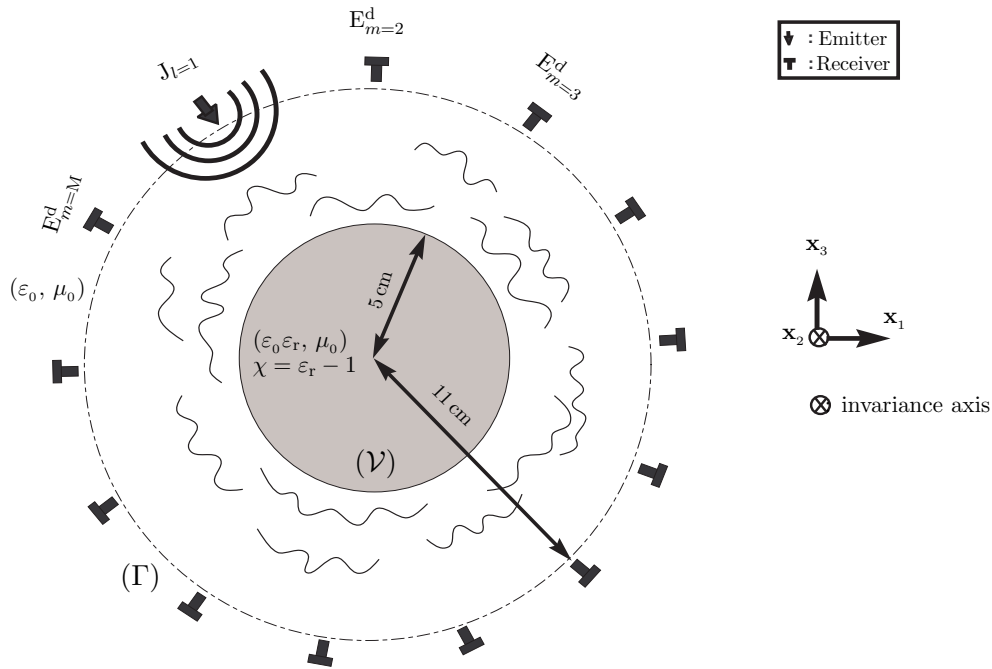} 
\caption{Configuration of the numerical study
\label{numerical-configuration}}
\end{psfrags}
\end{figure}

\begin{itemize}
\item[$\bullet$] \textbf{Configuration of the Study} In a fixed Cartesian coordinate system $({\rm O},\,{\mb x}_{_1},\,{\mb x}_{_2},\,{\mb x}_{_3})$, we consider a dielectric cylinder with circular cross section shape and with ${\rm O}{\mb x}_{_2}$ as its invariance axis Fig~(\ref{numerical-configuration}). This object is described by its permittivity-contrast function $\chi=\chi(x_{_1},\,x_{_3})$, defined in the domain of volume $(\mathcal{V})$. The latter is discretized with ${\rm Q}=3436$ square pixels of side $\Delta\simeq{1.5\,{\rm mm}}$ and area $\Delta\mathcal{V}\simeq{2.25\,{\rm mm}^{_2}}$. We are dealing with the case of ${\rm TE}$ polarization (electric component parallel to the invariance axis), and we model the electric component of the field, which is parallel to the invariance axis. The total field modelled in the object, and the scattered field simulated on the receivers, are respectively written: ${\mb E}={\rm E}(x_{_1},\,x_{_3})\,{\mb x}_{_2}$ and ${\mb E}^{\rm d}={\rm E}^{\rm d}(x_{_1},\,x_{_3})\,{\mb x}_{_2}$. A harmonic time dependence of the form ${\rm e}^{-{\rm i}\omega t}$ with pulsation $\omega$ is assumed. In this study, we consider ${\rm M}$ antennas arranged along a circle $(\Gamma)$ of radius $11\,{\rm cm}$, and concentric with the object of radius $5\,{\rm cm}$. All these antennas are situated at equal distance from the center of the object. Each one of them plays the role of emitter and receiver but not simultaneously. When an antenna $l$ acts like an emitter of source current ${\mb J}_l$ ($l\in\{1,\,2,...,\,{\rm M}\}$), all others act as receivers to detect the scattered field ${\mb E}^{\rm d}_m$, $m\in\{1,\,2,...,\,{\rm M}\}_{m\neq l}$. We work with a number of antennas ${\rm M}=16$, for a total of generated data equal to $240$.
\item[$\bullet$] \textbf{Incident Field} A source current ${\mb J}_l$ of unit amplitude ${\mb J}_l={\rm J}_l{\mb x}_{_2}=1\,{\mb x}_{_2}$, positioned at point ${\mb r}_l=x^{_l}_{_1}{\mb x}_{_1}+x^{_l}_{_3}{\mb x}_{_3}$ and oriented along the invariance axis ${\rm O}{\mb x}_{_2}$, radiates in two-dimensional free space the following incident field: 
\begin{equation}\label{incidentfield}
{\rm E}^{\rm i}_l({\mb r})\,{\mb x}_{_2}\,=\,-\frac{1}{4}\,\omega\mu_{_0}\,{\rm H}^{_{(1)}}_{_0}({\rm k}\,|{\mb r}-{\mb r}_l|)\,{\mb x}_{_2}
\end{equation}

Here, ${\rm E}^{\rm i}_l$ is obtained by solving the scalar Helmholtz equation \cite{abramowitz2012handbook}, and ${\rm k}=2\pi/\lambda$ denotes the wavevector modulus of the incident medium. ${\rm H}^{_{(1)}}_{_0}$ is the first-kind Hankel function of zero-order. The wavelength of the incident field is set to $\lambda=10\,{\rm cm}$ (i.e. 3 ${\rm GHz}$ of frequency).  

\end{itemize}

\subsection{Scattering Operator}
All vectorial physical quantities are aligned along the invariance axis. Therefore, the forward problem is recast as a scalar problem which is solved in the plane of incidence. This is done by the volume integral formulation, which builds a scattering operator. The latter consists of the following two coupled integral equations:

\begin{itemize}
\item[$\bullet$] \textbf{State Equation} Its solution is the total field ${\rm E}_l$. It is the result of an interaction between the incident field ${\rm E}^{\rm i}_l$ produced by the source current ${\mb J}_l$ and the object which is described by its permittivity-contrast function $\chi$.
\begin{equation}\label{etat}
{\rm E}_l({\mb r})\,=\,{\rm E}^{\rm i}_l({\mb r})\,+\,({\rm i}{\rm k}^{_2}/4)\int\limits_{\mathcal{V}}{\rm H}^{_{(1)}}_{_0}({\rm k}\,|{\mb r}-{\mb r}'|)\,\chi({\mb r}')\,{\rm E}_l({\mb r}')\,\,{\rm d}\mathcal{V}'\,,\,\,\,\,\,\,\,\,\,{\mb r}\in\mathcal{V}
\end{equation}
\item[$\bullet$] \textbf{Observation Equation} Once the total field ${\rm E}_l$ is computed using the state equation, it is replaced in the observation equation to model the diffracted field ${\rm E}_l^{\rm d}$ on the receiver, located at ${\mb r}_m$.
\begin{equation}\label{observation}
{\rm E}^{\rm d}_l({\mb r}_m)\,=\,({\rm i}{\rm k}^{_2}/4)\int\limits_{\mathcal{V}}{\rm H}^{_{(1)}}_{_0}({\rm k}\,|{\mb r}_m-{\mb r}'|)\,\chi({\mb r}')\,{\rm E}_l({\mb r}')\,\,{\rm d}\mathcal{V}'\,,\,\,\,\,\,\,\,\,\,\,\,\,\,\,\,\,\,\,\,\,\,\,\,\,{\mb r}_m\in\Gamma\,\text{ and}\,\,{\mb r}_m\neq {\mb r}_l
\end{equation}
\item[$\bullet$] \textbf{Symbolic Notation for the Scattering Operator and the Data} By varying the position ${\mb r}_l$ of the source current emitter ${\mb J}_l$, $l\in\{1,\,2,...,{\rm M}\}$, and the position ${\mb r}_m$, $m\in\{1,\,2,...,{\rm M}\}_{m\neq l}$ of the receiver, we generate data of the diffracted field, denoted: $\mathcal{D}=\{\mathcal{D}_{lm}={\rm E}^{\rm d}_l({\mb r}_m)\,\cdot\,{\mb J}_m\text{, for: }l,m\in\{1,\,2,...,\,{\rm M}\}_{l\neq m}\}$. The above equations Eqs.~(\ref{etat},\ref{observation}) give a nonlinear link between the data and the permittivity-contrast function, which from a mathematical point of view plays the role of a nonlinear scattering operator $\mathcal{N}$, such that: $\mathcal{D}=\mathcal{N}(\chi)$. Its numerical implementation is performed with the method of moments as described in references \cite{richmond1965scattering,lesselier1991buried,tsang2001scattering}.
\end{itemize} 

\subsection{Numerical Study}
\begin{itemize}
\item[$\bullet$] \textbf{Limited Functional Expansion of the Data} The functional derivatives expressed above can now be used to express the variation on the data $\delta\mathcal{D}=\{\delta\mathcal{D}_{lm}=\delta{\rm E}^{\rm d}_l({\mb r}_m)\,\cdot\,{\mb J}_m\text{, for : }l,m\in\{1,\,2,...,\,{\rm M}\}_{l\neq m}\}$ as a limited functional expansion with respect to the permittivity-contrast function $\chi({\mb r})$, for ${\mb r}\in\mathcal{V}$. We are interested in the convergence of this development which is written in the form: 

\begin{equation}
\delta\mathcal{D}\,=\,\sum\limits_{k=1}^{n}\mathcal{F}^{^{(k)}}(\underbrace{\delta\chi,\,...\,,\delta\chi}_{k\,\text{fois}})\,+\,{\rm o}(||\delta\chi||^{^{n}}) \\
\end{equation}

For odd $(n)$, each element $\delta\mathcal{D}_{lm}$ of $\delta\mathcal{D}$ is given by its expression in a condensed form:

\begin{equation}\label{DL7condensed}
\left\lbrace\begin{array}{l}
\delta\mathcal{D}_{lm}\,=\,\delta{\mb E}^{\rm d}_l({\mb r}_m)\,\cdot\,{\mb J}_m\,=\,\delta{\rm E}^{\rm d}_l({\mb r}_m) \\
\\
=\,\mathcal{F}_{_{lm}}^{^{(1)}}(\delta\chi)\,+\,\mathcal{F}_{_{lm}}^{^{(2)}}(\delta\chi,\delta\chi)\,+\,\mathcal{F}_{_{lm}}^{^{(3)}}(\delta\chi,\delta\chi,\delta\chi)\,+\,\mathcal{F}_{_{lm}}^{^{(4)}}(\underbrace{\delta\chi,\,...\,,\delta\chi}_{4\,\text{times}})\,+\,\mathcal{F}_{_{lm}}^{^{(5)}}(\underbrace{\delta\chi,\,...\,,\delta\chi}_{5\,\text{times}})\,+...\,\\
\\
\,\,\,\,\,\,\,\,\,\,\,\,\,\,\,\,\,\,\,\,\,\,\,\,\,\,\,\,\,\,\,\,\,\,\,\,\,\,+\mathcal{F}_{_{lm}}^{^{(2p)}}(\underbrace{\delta\chi,\,...\,,\delta\chi}_{2p\,\text{times}})\,+\,\mathcal{F}_{_{lm}}^{^{(2p+1)}}(\underbrace{\delta\chi,\,...\,,\delta\chi}_{2p+1\,\text{times}})\,+\,{\rm o}_{_{lm}}(||\delta\chi||^{^{2p+1}})
\end{array}\right.
\end{equation}
The explicit form of the above expansion, and the details of the numerical evaluation of the electric fields involved in its terms are described in appendix \ref{appendixc}.\\
\item[$\bullet$] \textbf{Convergence Criteria} To monitor the convergence of development Eq~(\ref{DL7condensed}), we introduce the residual error $\xi$ and study its behaviour as a function of the truncation order $(n)$:

\begin{equation}\label{ErrResDATA}
\left\lbrace\begin{array}{l}
\text{The residual error on data is the misfit: }\xi(n)\,=\,\Delta\mathcal{D}-\sum\limits_{k=1}^{n}\mathcal{F}^{^{(k)}}(\underbrace{\Delta\chi,\,...\,,\Delta\chi}_{k\,\text{times}})\\
\\
\text{$\mathcal{L}_{_2}$ norm of $\xi$ expressed as a percentage: }{\rm P}_{_\xi}(n)\,=\,100\,\times\,\sqrt{\frac{\sum\limits_{lm=1}^{\rm M^{^2}-M}|\xi_{_{lm}}(n)|^{^2}}{\sum\limits_{lm=1}^{\rm M^{^2}-M}|\Delta\mathcal{D}_{_{lm}}|^{^2}}} \\
\\
\text{$\mathcal{L}_{_2}$ norm of $\xi$ expressed in decibels: }{\rm L}_{_\xi}(n)\,=\,10\,\times\,\log_{_{10}}\,\left({\frac{\sum\limits_{lm=1}^{\rm M^{^2}-M}|\xi_{_{lm}}(n)|^{^2}}{\sum\limits_{lm=1}^{\rm M^{^2}-M}|\Delta\mathcal{D}_{_{lm}}|^{^2}}}\right) 
\end{array}\right.
\end{equation}

Note that the exact calculus of the difference on the data is directly obtained with the forward model, namely : $\Delta\mathcal{D}=\mathcal{N}(\chi+\Delta\chi)-\mathcal{N}(\chi)$. Concerning the variation on the contrast, it is numerically evaluated by the finite difference $\Delta\chi$. 
\vspace*{0.4cm}
\item[$\bullet$] \textbf{Numerical Results and Interpretation} We take as a numerical application, an object of homogeneous dielectric contrast $\chi_{_{\rm ref}}=6.0$. At this reference value, we add a difference $\Delta\chi$ giving a new object. Then, this finite difference introduced to modify the initial contrast is estimated with the expression ${\rm P}_{_{\chi}}=100\times(\Delta\chi/\chi_{_{\rm ref}})$. Figs~(\ref{fig1},\ref{fig2},\ref{fig3}) illustrate the convergence in decibels ${\rm L}_{_\xi}$ of the residual error $\xi$ with respect to the truncation order $n$, and for different values of ${\rm P}_{_{\chi}}$. For a better reading of these figures, we have added levels of value of this residual error, estimated with the expression of the percentage ${\rm P}_{_{\xi}}$. The latter are added to Figs~(\ref{fig1},\ref{fig2},\ref{fig3}), and are represented by horizontal dashed lines. In Fig~(\ref{fig1}), we consider the finite differences of the permittivity-contrast corresponding to the values ${\rm P}_{_{\chi}}=1\%,\,2\%,\,3\%,\,4\%\,\text{and}\,5\%$. For each case, we note that the convergence is reached for a truncation of order $n$, less than $15$. Moreover, we note that the speed of this convergence decreases with the increase of ${\rm P}_{_{\chi}}$. This tendency continues in Fig~(\ref{fig2}), for finite differences of the contrast corresponding to the values ${\rm P}_{_{\chi}}=6\%,\,7\%\,\text{et}\,8\%$. Indeed, for these three values, the convergence is obtained for $n<51$. For a finite difference of value ${\rm P}_{_{\chi}}=9\%$, we note a slow decrease in the residual error ${\rm L}_{_\xi}$, suggesting that the convergence occurs for a truncation order $n>51$. This is precisely what happens when the order of truncation is carried up to $n=161$ (see Fig~(\ref{fig3})). Concerning the value ${\rm P}_{_{\chi}}=10\%$, the increase of error ${\rm L}_{_\xi}$ according to the truncation order $n$ shown in Fig~(\ref{fig2}), is the mark that we are outside the domain of convergence of the functional limited expansion. These simulations were performed, with no particular optimisations, using MATLAB on a personal computer, with a quad-core processor at 2.60GHz and 16 Gb central memory. In particular, for a truncation order $n=161$, a simulation took $22$ minutes of computation time. \\
\begin{figure}[htpb]
\includegraphics[height=7cm,width=14cm]{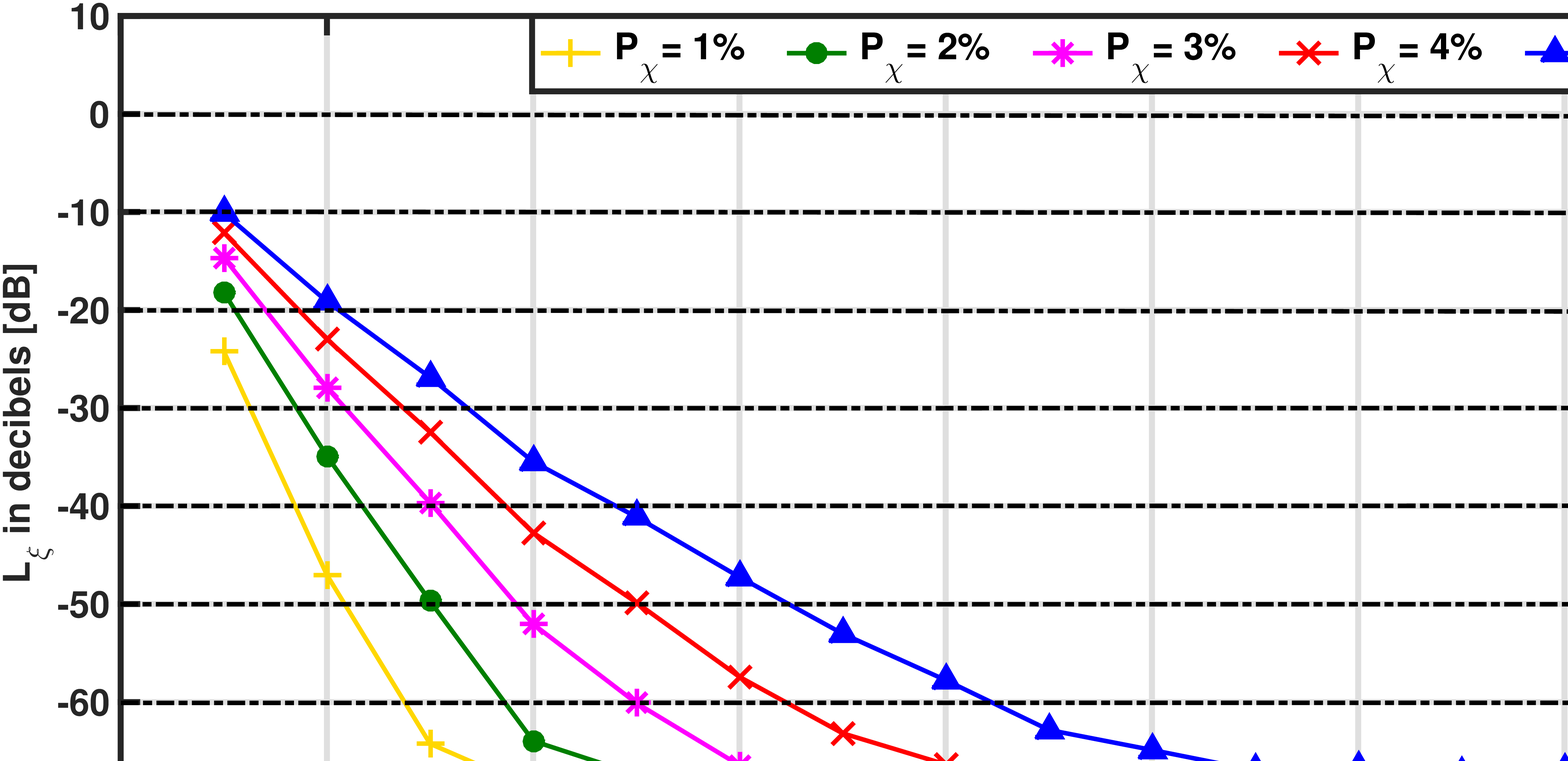} 
\vspace*{1.5cm}
\caption{Evolution of the residual error on data in decibels ${\rm L}_{_\xi}$ against the truncation order $n$ (with $1\leq n\leq 15$) of the limited functionnal expansion. Horizontal dashed lines give levels of value of this error calculated as a percentage ${\rm P}_{_{\xi}}$. The following finite differences of the contrast are considered : ${\rm P}_{_{\chi}}=1\%,\,2\%,\,3\%,\,4\%\,\text{and}\,5\%$\label{fig1}}
\end{figure}
\begin{figure}[htpb]
\includegraphics[height=7cm,width=14cm]{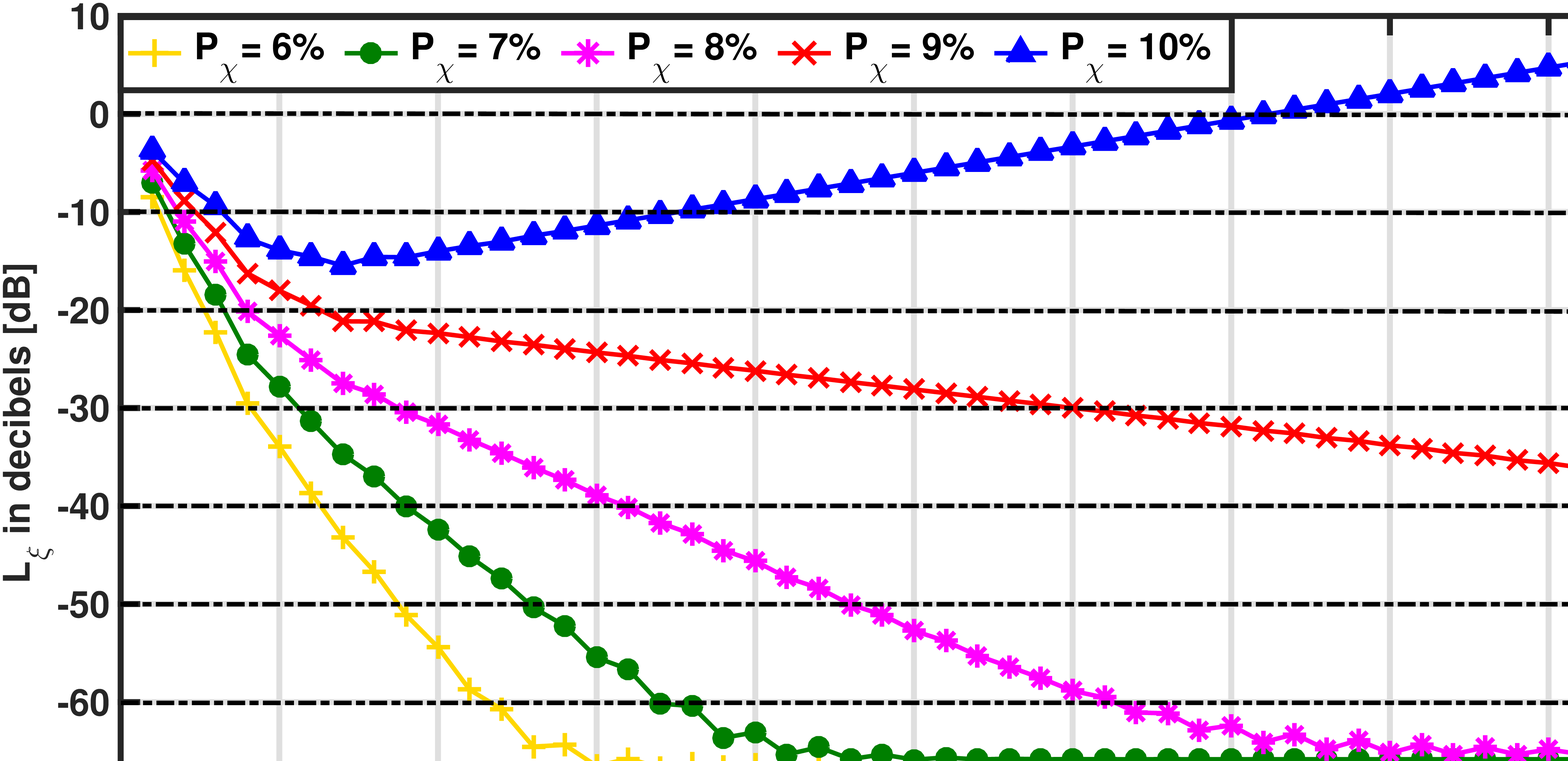} 
\vspace*{1.5cm}
\caption{Evolution of the residual error on data in decibels ${\rm L}_{_\xi}$ against the truncation order $n$ (with $1\leq n\leq 51$) of the limited functionnal expansion. Horizontal dashed lines give levels of value of this error calculated as a percentage ${\rm P}_{_{\xi}}$. The following finite differences of the contrast are considered : ${\rm P}_{_{\chi}}=6\%,\,7\%,\,8\%,\,9\%\,\text{and}\,10\%$\label{fig2}}
\end{figure}
\begin{figure}[htpb]
\includegraphics[height=7cm,width=14cm]{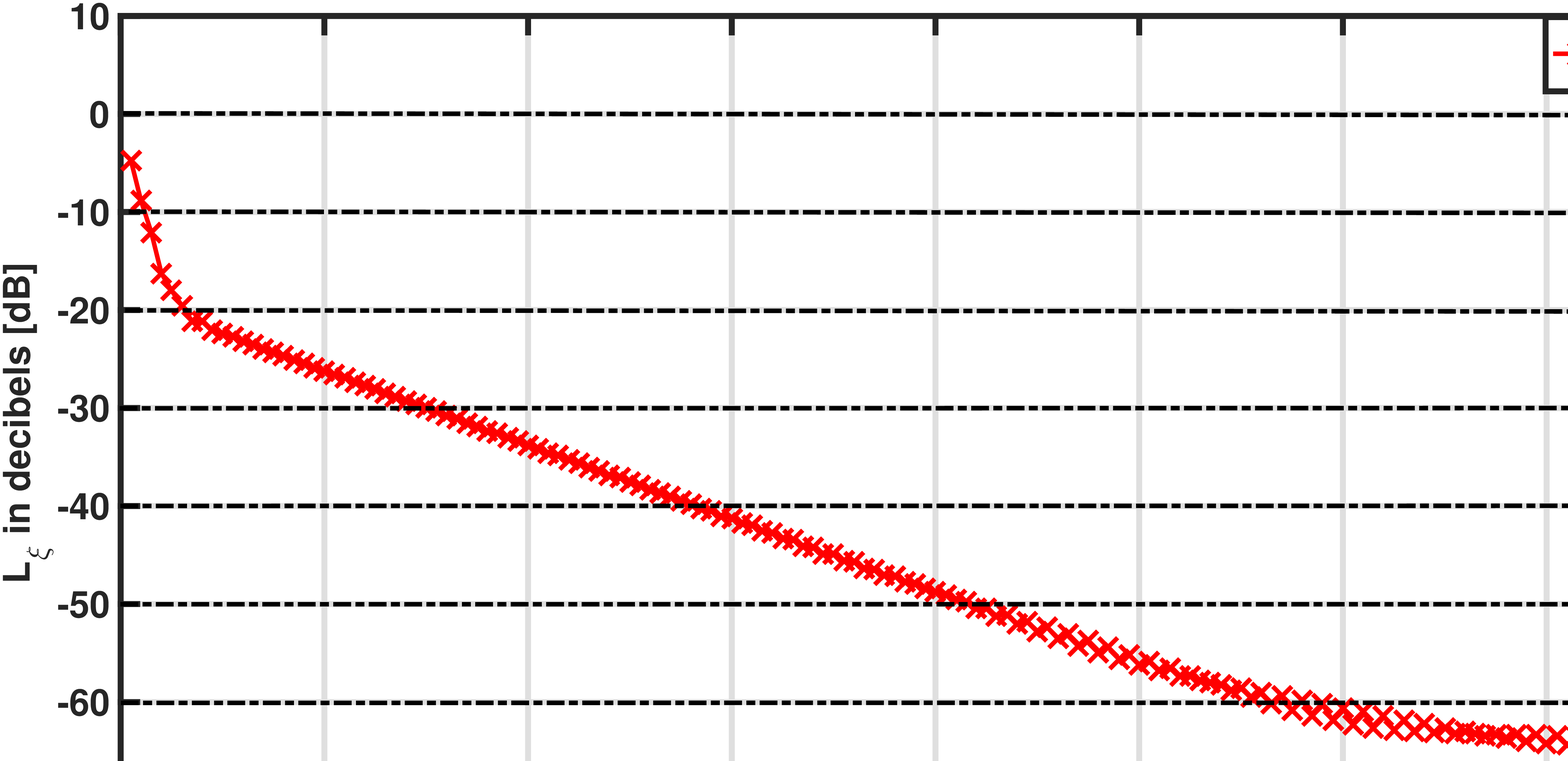} 
\vspace*{1.5cm}
\caption{Evolution of the residual error on data in decibels ${\rm L}_{_\xi}$ against the truncation order $n$ (with $1\leq n\leq 161$) of the limited functionnal expansion. Horizontal dashed lines give levels of value of this error calculated as a percentage ${\rm P}_{_{\xi}}$. The following finite difference of the contrast is considered : ${\rm P}_{_{\chi}}=9\%$\label{fig3}}
\end{figure}
\end{itemize}
\section{Conclusion}
\label{sec:Conclusion}
Previous studies have shown that use of an approach involving the reciprocity theorem, combined with a formulation of Maxwell's equations in the sense of distributions, is an efficient approach to calculate the Fr\'echet derivative. The latter stands for the functional derivative of the first order of the diffracted field, with respect to the permittivity-contrast function. The remainder, which contains all the derivatives of higher orders, has been neglected. In the present work, we have applied this principle of calculus on the remainder, and have succeeded in obtaining all the functional derivatives of higher orders. This allowed us to express, the variation of the data as a limited functional expansion with respect to the permittivity-contrast function. Finally, we have numerically illustrated the convergence of this development, in the case of a diffraction by a dielectric cylindrical object illuminated by an incident field in ${\rm TE}$ polarization. Such an expansion can be used to propose inversion methods in research areas involving the problem of permittivity-contrast reconstruction. We expect some improvements in the quality of the solutions as well as the speed of convergence.
\appendixx{}
\label{appendixa}
Let's start with the equality:
\begin{equation}\label{Distributionprime}
\nabla\,\cdot\,[{\mb H}_{{\rm b},m}\,\times\,{\mb E}_{{\rm a},l}\,-\,{\mb H}_{{\rm a},l}\,\times\,{\mb E}_{{\rm b},m}]\,=\,{-\rm i}\,\omega\,(\varepsilon_{\rm b}\,-\,\varepsilon_{\rm a})\,{\mb E}_{{\rm a},l}\,\cdot\,{\mb E}_{{\rm b},m}\,+\,\delta_{{\mb r}_m}\,{\mb E}_{{\rm a},l}\,\cdot\,{\mb J}_m\,-\,\delta_{{\mb r}_l}\,{\mb E}_{{\rm b},m}\,\cdot\,{\mb J}_l
\end{equation} 
 
Each term of the above equation being a distribution, we can apply it on a test function. Such a function is denoted $u({\mb r})$ and is chosen so as to satisfy the following conditions:

\begin{itemize}
\item[$\bullet$] $u({\mb r})$ is a function with bounded support.
\item[$\bullet$] $u({\mb r})$ is a function of class ${\mt C}^\infty$.
\end{itemize}

We add the following additional conditions without loss of generality \cite{appel2007}: 

\begin{equation}
\left\lbrace\begin{array}{ccccc}           
u({\mb r})\,=\,1 & {\mb r}\,\in\,{\rm C}_{\rm i} &  \text{and} &  u({\mb r})\,=\,1 & {\mb r}\,\in\,{\rm C} \\ 
\nabla u({\mb r})\,\neq\,0 & {\mb r}\,\in\,{\rm F}_{\rm i} &  \text{and} &  u({\mb r})\,=\,0 & {\mb r}\,\in\,{\rm E}  
\end{array}\right.
\end{equation}

Here, ${\rm C}$ denotes the surface of a sphere of volume ${\rm C}_{\rm i}$, ${\rm E}$ the surface of a larger sphere including the first and having the volume ${\rm E}_{\rm i}$. The term ${\rm F}_{\rm i}\,=\,{\rm E}_{\rm i}\,-\,{\rm C}_{\rm i}$ corresponds to the volume comprised between the surfaces ${\rm C}$ and ${\rm E}$. In practice, it is possible to construct a test function that satisfies these three criteria. We proceed now on the application of each term of the Eq~(\ref{Distributionprime}) to the test function $u({\mb r})$. Starting with its left-hand side, setting ${\mb T}\,=\,{\mb H}_{{\rm b},m}\,\times\,{\mb E}_{{\rm a},l}\,-\,{\mb H}_{{\rm a},l}\,\times\,{\mb E}_{{\rm b},m}$, and applying then the distribution $\nabla\,\cdot\,{\mb T}$ on the test function $u$, it follows:

\begin{equation}
<\nabla\,\cdot\,{\mb T}|u>\,=\,-<{\mb T}|\nabla u>\,=\,-\int\limits_{{\rm F}_{\rm i}}\,{\mb T}\,\cdot\,\nabla u\,\,{\rm d}{\rm F}_{\rm i}\,=\,-\int\limits_{{\rm F}_{\rm i}}\,\nabla\,\cdot\,({\mb T}u)\,\,{\rm d}{\rm F}_{\rm i}\,+\int\limits_{{\rm F}_{\rm i}}\,(\nabla\,\cdot\,{\mb T})\,u\,\,{\rm d}{\rm F}_{\rm i}
\end{equation}

The integral $\int\limits_{{\rm F}_{\rm i}}\,(\nabla\,\cdot\,{\mb T})\,u\,\,{\rm d}{\rm F}_{\rm i}$ is nil because, according to Equation Eq~(\ref{Distributionprime}), each term of its right-hand side is equal to zero for all ${\mb r}\in{\rm F}_{\rm i}$. On the other hand, the divergence theorem allows us to transform the volume integral $-\int\limits_{{\rm F}_{\rm i}}\,\nabla\,\cdot\,({\mb T}u)\,\,{\rm d}{\rm F}_{\rm i}$ into surface integral:

\begin{equation}
<\nabla\,\cdot\,{\mb T}|u>\,=\,-\int\limits_{{\rm F}_{\rm i}}\,\nabla\,\cdot\,({\mb T}\,u)\,\,{\rm d}{\rm F}_{\rm i}\,=\,+\int\limits_{\rm C}\,{\mb T}u\,\cdot\,{\mb e}_{_{\rho}}\,\,{\rm d}{\rm C}\,-\int\limits_{\rm E}\,{\mb T}u\,\cdot\,{\mb e}_{_{\rho}}\,\,{\rm d}{\rm E}
\end{equation}

Here ${\mb e}_{_{\rho}}$ denotes the unit vector orthogonal to the surfaces ${\rm C}$ and ${\rm E}$. From the third property of the test function $u$, we have : $u({\mb r})=0$ $\forall\,{\mb r}\in{\rm E}$ and $u({\mb r})=1$ $\forall\,{\mb r}\in{\rm C}$. This allows us to write the following result for the calculus of the left-hand side of equation Eq~(\ref{Distributionprime}): 

\begin{equation}\label{Membreg}
<\nabla\,\cdot\,{\mb T}|u>\,=\,+\int\limits_{\rm C}\,{\mb T}\,\cdot\,{\mb e}_{_{\rho}}\,\,{\rm d}{\rm C}\,=\,+\int\limits_{\rm C}\,[{\mb H}_{{\rm b},m}\,\times\,{\mb E}_{{\rm a},l}\,-\,{\mb H}_{{\rm a},l}\,\times\,{\mb E}_{{\rm b},m}]\,\cdot\,{\mb e}_{_{\rho}}\,\,{\rm d}{\rm C}
\end{equation}

Let us now apply the right-hand side of equation Eq~(\ref{Distributionprime}) on the test function $u$. This gives directly:

\begin{equation}\label{Membred}
\begin{array}{l}
<{-\rm i}\,\omega\,(\varepsilon_{\rm b}\,-\,\varepsilon_{\rm a})\,{\mb E}_{{\rm a},l}\,\cdot\,{\mb E}_{{\rm b},m}\,+\,\delta_{{\mb r}_m}\,{\mb E}_{{\rm a},l}\,\cdot\,{\mb J}_m\,-\,\delta_{{\mb r}_l}\,{\mb E}_{{\rm b},m}\,\cdot\,{\mb J}_l|u>\,=\, \\
\\
\,\,\,\,\,\,\,\,\,\,\,\,\,\,\,\,\,\,\,\,\,\,\,\,\,\,\,\,\,\,\,\,\,\,\,\,{-\rm i}\,\omega\int\limits_{\mathcal{V}}\,(\varepsilon_{\rm b}\,-\,\varepsilon_{\rm a})\,{\mb E}_{{\rm a},l}\,\cdot\,{\mb E}_{{\rm b},m}\,\,{\rm d}\mathcal{V}\,+\,{\mb E}_{{\rm a},l}({\mb r}_m)\,\cdot\,{\mb J}_m\,-\,{\mb E}_{{\rm b},m}({\mb r}_l)\,\cdot\,{\mb J}_l
\end{array}
\end{equation}
 
Now, in view of Eqs~(\ref{Membreg},\ref{Membred}), we can rewrite Eq~(\ref{Distributionprime}) in the sense of functions, in the form:

\begin{equation}\label{Fonction1prime}
\begin{array}{l}
\int\limits_{\rm C}\,[{\mb H}_{{\rm b},m}\,\times\,{\mb E}_{{\rm a},l}\,-\,{\mb H}_{{\rm a},l}\,\times\,{\mb E}_{{\rm b},m}]\,\cdot\,{\mb e}_{_{\rho}}\,\,{\rm d}{\rm C}\,=\,{-\rm i}\,\omega\int\limits_{\mathcal{V}}\,(\varepsilon_{\rm b}\,-\,\varepsilon_{\rm a})\,{\mb E}_{{\rm a},l}\,\cdot\,{\mb E}_{{\rm b},m}\,\,{\rm d}\mathcal{V}\,+\,{\mb E}_{{\rm a},l}({\mb r}_m)\,\cdot\,{\mb J}_m\,-\, \\
\,\,\,\,\,\,\,\,\,\,\,\,\,\,\,\,\,\,\,\,\,\,\,\,\,\,\,\,\,\,\,\,\,\,\,\,\,\,\,\,\,\,\,\,\,\,\,\,\,\,\,\,\,\,\,\,\,\,\,\,\,\,\,\,\,\,\,\,\,\,\,\,\,\,\,\,\,\,\,\,\,\,\,\,\,\,\,\,\,\,\,\,\,\,\,\,\,\,\,\,\,\,\,\,\,\,\,\,\,\,\,\,\,\,\,{\mb E}_{{\rm b},m}({\mb r}_l)\,\cdot\,{\mb J}_l
\end{array}
\end{equation}

\appendixx{}
\label{appendixb}

To show that the integral $\int\limits_{{\rm C}}\,[{\mb H}_{{\rm b},m}\,\times\,{\mb E}_{{\rm a},l}\,-\,{\mb H}_{{\rm a},l}\,\times\,{\mb E}_{{\rm b},m}]\,\cdot\,{\mb e}_{_{\rho}}\,\,{\rm d}{\rm C}$ is nil, let ${\rm C}\,=\,{\rm C}_{\infty}$ be the surface of a sphere with a radius large enough to be in the farfield region. If we consider the configuration (${\rm a}_{_l}$), its total magnetic field ${\mb H}_{{\rm a},l}$ is given by the sum between the diffracted field, and the incident field radiated by a point-like source current located at a finite distance from the considered object. In the farfield region, this total field satisfies the Sommerfeld radiation condition $\lim\limits_{_{_{\rho\rightarrow +\infty}}}\rho\,[\frac{\partial}{\partial\rho}-{\rm i}k]\,{\mb H}_{{\rm a},l}=0$, and is written as a product of a radial function by an angular function ${\mb H}_{{\rm a},l}\,=\,\frac{e^{{\rm i}k\rho}}{4\pi\rho}\,{\mb A}_{{\rm a},l}(\theta,\phi)$, with both expressed in spherical coordinates \cite{yaghjian1986overview,tsang2000scattering}. The magnetic farfields of the configurations (${\rm a}_{_l}$) and (${\rm b}_{_m}$) are given by the following expressions:

\begin{equation}\label{farone}
\left\lbrace\begin{array}{l}   
{\mb H}_{{\rm a},l}\,=\,\frac{e^{{\rm i}k\rho}}{4\pi\rho}\,{\mb A}_{{\rm a},l}\,\propto\,\frac{e^{{\rm i}k\rho}}{\rho}\,{\mb A}_{{\rm a},l} \\
\\
{\mb H}_{{\rm b},m}\,=\,\frac{e^{{\rm i}k\rho}}{4\pi\rho}\,{\mb A}_{{\rm b},m}\,\propto\,\frac{e^{{\rm i}k\rho}}{\rho}\,{\mb A}_{{\rm b},m}
\end{array}\right.
\end{equation}

Due to the relation ${\mb E}=\frac{1}{\omega\varepsilon_{_0}}\nabla\times{\mb H}$, the electric field is obtained from the magnetic field. After a few calculations, we have:

\begin{equation}\label{fartwo}
\left\lbrace\begin{array}{l}   
{\mb E}_{{\rm a},l}\,=\,\frac{{\rm i}k}{4\pi\omega\varepsilon_{_0}}\,\frac{e^{{\rm i}k\rho}}{\rho}\,{\mb e}_{_{\rho}}\,\times\,{\mb A}_{{\rm a},l}\,\propto\,\frac{e^{{\rm i}k\rho}}{\rho}\,{\mb e}_{_{\rho}}\,\times\,{\mb A}_{{\rm a},l} \\
\\
{\mb E}_{{\rm b},m}\,=\,\frac{{\rm i}k}{4\pi\omega\varepsilon_{_0}}\,\frac{e^{{\rm i}k\rho}}{\rho}\,{\mb e}_{_{\rho}}\,\times\,{\mb A}_{{\rm b},m}\,\propto\,\frac{e^{{\rm i}k\rho}}{\rho}\,{\mb e}_{_{\rho}}\,\times\,{\mb A}_{{\rm b},m}
\end{array}\right.
\end{equation}

Here ${\mb e}_{_{\rho}}$ denotes the radial unit vector. In virtue of expressions Eqs~(\ref{farone},\ref{fartwo}), the following integral vanishes:

\begin{equation}
\left\lbrace\begin{array}{l}           
\int\limits_{{\rm C_\infty}}\,[{\mb H}_{{\rm b},m}\,\times\,{\mb E}_{{\rm a},l}\,-\,{\mb H}_{{\rm a},l}\,\times\,{\mb E}_{{\rm b},m}]\,\cdot\,{\mb e}_{_{\rho}}\,\,{\rm d}{\rm C}\\
\\
=\frac{e^{{\rm i}2k\rho}}{\rho^2}\int\limits_{{\rm C}_\infty}\,[\{({\mb A}_{{\rm b},m}\,\cdot\,{\mb A}_{{\rm a},l})\,({\mb e}_{_{\rho}}\,\cdot\,{\mb e}_{_{\rho}})\,-\,({\mb A}_{{\rm b},m}\,\cdot\,{\mb e}_{_{\rho}})\,({\mb A}_{{\rm a},l}\,\cdot\,{\mb e}_{_{\rho}})\}\,-\, \\
\,\,\,\,\,\,\,\,\,\,\,\,\,\,\,\,\,\,\,\,\,\,\,\,\,\,\,\,\,\,\,\,\,\,\,\,\,\,\,\,\,\,\,\,\,\,\,\,\,\,\,\,\,\,\,\,\,\,\,\,\,\,\,\,\,\,\,\,\,\,\,\,\{({\mb A}_{{\rm a},l}\,\cdot\,{\mb A}_{{\rm b},m})({\mb e}_{_{\rho}}\,\cdot\,{\mb e}_{_{\rho}})\,-\,({\mb A}_{{\rm a},l}\,\cdot\,{\mb e}_{_{\rho}})\,({\mb A}_{{\rm b},m}\,\cdot\,{\mb e}_{_{\rho}})\}]\,\,{\rm d}{{\rm C}}=0
\end{array}\right.
\end{equation}

\appendixx{}
\label{appendixc}

For odd $(n)$, each element $\delta\mathcal{D}_{lm}$ of $\delta\mathcal{D}$ is given by the following detailed expression:

\begin{equation}\label{DL7plus}
\begin{array}{l}
\delta\mathcal{D}_{lm}\,=\,\delta{\mb E}^{\rm d}_l({\mb r}_m)\,\cdot\,{\mb J}_m\,=\,\delta{\rm E}^{\rm d}_l({\mb r}_m) \\
\\
=\,\underbrace{{-\rm i}\omega\varepsilon_{_0}\int\limits_{\mathcal{V}}\,\delta\chi({\mb r})\,\,{\rm E}_l({\mb r})\,\,{\rm E}_m({\mb r})\,{\rm d}\mathcal{V}}_{\mathcal{F}_{_{lm}}^{^{(1)}}(\delta\chi)}\,\,\underbrace{{-\rm i}\omega\varepsilon_{_0}\int\limits_{\mathcal{V}}\,\delta\chi({\mb r})\,\,{\rm E}_m({\mb r})\,\,{\rm E}_{l,{\mb r}^{*}_{\scalebox{0.4}{(1)}}}({\mb r})\,{\rm d}\mathcal{V}}_{\mathcal{F}_{_{lm}}^{^{(2)}}(\delta\chi,\delta\chi)} \\
\\
\,\,\,\,\,\,\,\underbrace{{-\rm i}\omega\varepsilon_{_0}\int\limits_{\mathcal{V}}\,\delta\chi({\mb r})\,\,{\rm E}_{l,{\mb r}^{*}_{\scalebox{0.4}{(1)}}}({\mb r})\,\,{\rm E}_{m,{\mb r}^{*}_{\scalebox{0.4}{(1)}}}({\mb r})\,{\rm d}\mathcal{V}}_{\mathcal{F}_{_{lm}}^{^{(3)}}(\delta\chi,\delta\chi,\delta\chi)}\,\,\underbrace{{-\rm i}\omega\varepsilon_{_0}\int\limits_{\mathcal{V}}\,\delta\chi({\mb r})\,\,{\rm E}_{m,{\mb r}^{*}_{\scalebox{0.4}{(1)}}}({\mb r})\,\,{\rm E}_{l,{\mb r}^{*}_{\scalebox{0.4}{(2)}}}({\mb r})\,{\rm d}\mathcal{V}}_{\mathcal{F}_{_{lm}}^{^{(4)}}(\underbrace{\delta\chi,\,...\,,\delta\chi}_{4\,\text{times}})} \\
\\
\,\,\,\,\,\,\,\underbrace{{-\rm i}\omega\varepsilon_{_0}\int\limits_{\mathcal{V}}\,\delta\chi({\mb r})\,\,{\rm E}_{l,{\mb r}^{*}_{\scalebox{0.4}{(2)}}}({\mb r})\,\,{\rm E}_{m,{\mb r}^{*}_{\scalebox{0.4}{(2)}}}({\mb r})\,{\rm d}\mathcal{V}}_{\mathcal{F}_{_{lm}}^{^{(5)}}(\underbrace{\delta\chi,\,...\,,\delta\chi}_{5\,\text{times}})}\,\,...\,+\,...\,+\,...\,\,\underbrace{{-\rm i}\omega\varepsilon_{_0}\int\limits_{\mathcal{V}}\,\delta\chi({\mb r})\,\,{\rm E}_{m,{\mb r}^{*}_{\scalebox{0.4}{(p-1)}}}({\mb r})\,\,{\rm E}_{l,{\mb r}^{*}_{\scalebox{0.4}{(p)}}}({\mb r})\,\,{\rm d}\mathcal{V}}_{\mathcal{F}_{_{lm}}^{^{(2p)}}(\underbrace{\delta\chi,\,...\,,\delta\chi}_{2p\,\text{times}})} \\
\\
\,\,\,\,\,\,\,\underbrace{{-\rm i}\omega\varepsilon_{_0}\int\limits_{\mathcal{V}}\,\delta\chi({\mb r})\,\,{\rm E}_{l,{\mb r}^{*}_{\scalebox{0.4}{(p)}}}({\mb r})\,\,{\rm E}_{m,{\mb r}^{*}_{\scalebox{0.4}{(p)}}}({\mb r})\,\,{\rm d}\mathcal{V}}_{\mathcal{F}_{_{lm}}^{^{(2p+1)}}(\underbrace{\delta\chi,\,...\,,\delta\chi}_{2p+1\,\text{times}})}\,\,\underbrace{{-\rm i}\omega\varepsilon_{_0}\int\limits_{\mathcal{V}}\,\delta\chi({\mb r})\,\,{\rm E}_{l,{\mb r}^{*}_{\scalebox{0.4}{(p)}}}({\mb r})\,\,\delta{\rm E}_{m,{\mb r}^{*}_{\scalebox{0.4}{(p)}}}({\mb r})\,\,{\rm d}\mathcal{V}}_{{\rm o}_{_{lm}}(||\delta\chi||^{^{2p+1}})}
\end{array}
\end{equation}

The following steps describe how to calculate the total fields ${\rm E}_w({\mb r})$, ${\rm E}_{w,{\mb r}^{*}_{\scalebox{0.4}{(1)}}}({\mb r})$, ${\rm E}_{w,{\mb r}^{*}_{\scalebox{0.4}{(2)}}}({\mb r})$,...,\,${\rm E}_{w,{\mb r}^{*}_{\scalebox{0.4}{(p)}}}({\mb r})$ (with $w=l,m$):

\begin{itemize}
\vspace*{0.5cm}
\item[$\bullet$] \textbf{\underline{Step 0}:} In view of the Eq~(\ref{incidentfield}), the source current ${\rm J}_w=1$ located at ${\mb r}_w$ radiates the incident field ${\rm E}^{\rm i}_w({\mb r})\,=\,-\frac{1}{4}\,\omega\mu_{_0}\,{\rm H}^{_{(1)}}_{_0}({\rm k}\,|{\mb r}-{\mb r}_w|)$, (${\mb r}\in\mathcal{V}$). The resulting total field, solution of the state equation Eq~(\ref{etat}) is denoted by ${\rm E}_w({\mb r})$, (${\mb r}\in\mathcal{V}$).
\vspace*{0.5cm}
\item[$\bullet$] \textbf{\underline{Step 1}:} The domain $\mathcal{V}$ is meshed with ${\rm Q}$ pixels of area $\Delta\mathcal{V}$ centred on the positions ${\mb r}_q$ (with $q=1,2,...,{\rm Q}$), the source currents $-{\rm i}\omega\varepsilon_{\rm o}\,\Delta\chi({\mb r}_q)\,{\rm E}_w({\mb r}_q)\,\Delta\mathcal{V}$ (with ${\mb r}_q$ spanning the domain $\mathcal{V}$), radiate the total incident field ${\rm E}^{\rm i}_{w,{\mb r}^{*}_{\scalebox{0.4}{(1)}}}({\mb r})={\rm i}({\rm k}^{_2}/4)\,\Delta\mathcal{V}\sum\limits_{q=1}^{\rm Q}\Delta\chi({\mb r}_q)\,{\rm E}_w({\mb r}_q)\,{\rm H}^{_{(1)}}_{_0}({\rm k}\,|{\mb r}-{\mb r}_q|)$, (${\mb r},{\mb r}_q\in\mathcal{V}$). Note that when ${\mb r}={\mb r}_q$, the value of ${\rm H}^{_{(1)}}_{_0}$ is obtained by performing the calculus of the average $(1/\Delta\mathcal{V})\int\limits_{\Delta\mathcal{V}}{\rm H}^{_{(1)}}_{_0}({\rm k}\,|{\mb r}-{\mb r}_q|){\rm d}{\mb r}$, where $\Delta\mathcal{V}$ is the area of the pixel $q$ centred on the position ${\mb r}_q$ (details on this calculus are presented in \cite{lesselier1991buried}). As stated in section~\ref{sec:Theoretical Approach}, the index ${\mb r}^{*}_{\scalebox{0.4}{(1)}}$ has two significations: the symbol $*$ means that we sum over all positions ${\mb r}_q$ ($q=1,2,...,{\rm Q}$), in order to take into account the contributions of all the source currents in the total incident field. The index $(1)$ means that the variation of the permittivity-contrast function $\delta\chi$ (replaced in this numerical study by the finite difference $\Delta\chi$), is included once in the expression of the total incident field ${\rm E}^{\rm i}_{w,{\mb r}^{*}_{\scalebox{0.4}{(1)}}}$. Then, the total electric field ${\rm E}_{w,{\mb r}^{*}_{\scalebox{0.4}{(1)}}}({\mb r})$, (${\mb r}\in\mathcal{V}$) is obtained by solving the state equation Eq~(\ref{etat}), with ${\rm E}^{\rm i}_{w,{\mb r}^{*}_{\scalebox{0.4}{(1)}}}$, (${\mb r}\in\mathcal{V}$) playing the role of incident field.
\vspace*{0.5cm}
\item[$\bullet$] \textbf{\underline{Step 2}:} In this step we replace ${\rm E}_w$ by ${\rm E}_{w,{\mb r}^{*}_{\scalebox{0.4}{(1)}}}$, and we repeat the same calculus as in the previous step, namely: all source currents $-{\rm i}\omega\varepsilon_{\rm o}\,\Delta\chi({\mb r}_q)\,{\rm E}_{w,{\mb r}^{*}_{\scalebox{0.4}{(1)}}}({\mb r}_q)\,\Delta\mathcal{V}$ (with ${\mb r}_q$ spanning the domain $\mathcal{V}$), radiate the total incident field ${\rm E}^{\rm i}_{w,{\mb r}^{*}_{\scalebox{0.4}{(2)}}}({\mb r})={\rm i}({\rm k}^{_2}/4)\,\Delta\mathcal{V}\sum\limits_{q=1}^{\rm Q}\Delta\chi({\mb r}_q)\,{\rm E}_{w,{\mb r}^{*}_{\scalebox{0.4}{(1)}}}({\mb r}_q)\,{\rm H}^{_{(1)}}_{_0}({\rm k}\,|{\mb r}-{\mb r}_q|)$, (${\mb r},{\mb r}_q\in\mathcal{V}$). The index $(2)$ means that the finite difference $\Delta\chi$ is included twice in the expression of the total incident field ${\rm E}^{\rm i}_{w,{\mb r}^{*}_{\scalebox{0.4}{(2)}}}$ (once explicitly and once implicitly because embedded in the field ${\rm E}_{w,{\mb r}^{*}_{\scalebox{0.4}{(1)}}}$). Finally, the total field ${\rm E}_{w,{\mb r}^{*}_{\scalebox{0.4}{(2)}}}({\mb r})$, (${\mb r}\in\mathcal{V}$) is obtained in the same way, that is to say by solving the state equation Eq~(\ref{etat}), with the term ${\rm E}^{\rm i}_{w,{\mb r}^{*}_{\scalebox{0.4}{(2)}}}({\mb r})$, (${\mb r}\in\mathcal{V}$) as the incident field.\\
.........................\\
.........................\\
.........................\\
\item[$\bullet$] \textbf{\underline{Step p}:} The source currents $-{\rm i}\omega\varepsilon_{\rm o}\,\Delta\chi({\mb r}_q)\,{\rm E}_{w,{\mb r}^{*}_{\scalebox{0.4}{(p-1)}}}({\mb r}_q)\,\Delta\mathcal{V}$ (with ${\mb r}_q$ spanning the domain $\mathcal{V}$), radiate the total incident field ${\rm E}^{\rm i}_{w,{\mb r}^{*}_{\scalebox{0.4}{(p)}}}({\mb r})={\rm i}({\rm k}^{_2}/4)\,\Delta\mathcal{V}\sum\limits_{q=1}^{\rm Q}\Delta\chi({\mb r}_q)\,{\rm E}_{w,{\mb r}^{*}_{\scalebox{0.4}{(p-1)}}}({\mb r}_q)\,{\rm H}^{_{(1)}}_{_0}({\rm k}\,|{\mb r}-{\mb r}_q|)$, (${\mb r},{\mb r}_q\in\mathcal{V}$). The index $(p)$ means that the finite difference $\Delta\chi$ is included $p$ times in the expression of the total incident field ${\rm E}^{\rm i}_{w,{\mb r}^{*}_{\scalebox{0.4}{(p)}}}$ (once explicitly and $p-1$ times implicitly because embedded in the field ${\rm E}_{w,{\mb r}^{*}_{\scalebox{0.4}{(p-1)}}}$). The total field ${\rm E}_{w,{\mb r}^{*}_{\scalebox{0.4}{(p)}}}({\mb r})$, (${\mb r}\in\mathcal{V}$) is then obtained by solving the state equation Eq~(\ref{etat}), for which ${\rm E}^{\rm i}_{w,{\mb r}^{*}_{\scalebox{0.4}{(p)}}}({\mb r})$, (${\mb r}\in\mathcal{V}$) is used as the incident field.\\
\end{itemize}

\end{document}